\documentclass[aps,prc,twocolumn,showpacs,preprintnumbers,amsmath,amssymb,superscriptaddress]{revtex4}
\usepackage{graphicx}				
\usepackage{amssymb}
\usepackage{dcolumn}
\usepackage{bm}
\newcommand{ \la }{\langle}

\newcommand{ \ra }{\rangle}

\def \mean#1 {{\la #1 \ra}}
\def \etal {{ \it et al.\ }}

\begin{document}
\title{Correcting Correlation Function Measurements}
\author{Shantam Ravan, Prabhat Pujahari, Sidharth Prasad, and Claude A. Pruneau\\
Department of Physics and Astronomy, Wayne State University}
\date{\today}
\begin{abstract}
Correlation functions measured as a function of $\Delta \eta, \Delta \phi$ have emerged as a powerful tool to study the dynamics of particle production in nuclear collisions at high energy. They are however subject, like any other observables, to instrumental effects which must be properly accounted for to extract meaningful physics results. We compare the merits of several techniques used towards  measurement of these correlation functions  in 
nuclear collisions. We discuss and distinguish the effects of finite acceptance, and detection efficiency that may vary with collision parameters such
as the position of the event in the detector and the instantaneous luminosity of the beam. We focus in particular on 
instrumental effects which break the factorization of the particle pair detection efficiency, and describe a technique to recover the
robustness of correlation observables.  We finally introduce a multi-dimensional
 weight method  to correct for efficiencies that vary simultaneously with particle pseudo rapidity, azimuthal angle, transverse momentum, and the 
 collision vertex position. The method can be generalized to account for any number of "event variables" that may break the factorability of the
 pair efficiency.
\end{abstract}
\maketitle

\setcounter{page}{1}

\section{Introduction}

Over the last decade and half, correlation functions have emerged as one of the best tools to characterize particle production and study the dynamics of heavy ion collisions. And yet, recent postings on public archives suggest several misconceptions associated with measurements, analysis, and interpretation of correlation functions may still exist. Indeed, a number of talks and notes claim event mixing is wrong, and that new techniques must be invented to correct raw correlation data~\cite{Xu2013}.   In this paper, we endeavor to clarify the notion of correlation function, the impact of the detector acceptance, as well as fake correlations introduced by various instrumental effects. Instrumental effects can be grouped in distinct categories. These include the effect of the finite measurement acceptance, detection efficiencies, variation of the acceptance or efficiency with ÒexternalÓ variables such as the collision vertex position, the detector occupancy and beam luminosity, and other  detector conditions that can influence the overall performance of the detector. We  discuss  several of these effects and study their relative importance with an illustrative correlation model. We also describe  techniques to remedy or at the very least mitigate these effects. Much of what is presented here is not new. However, the acceptance and efficiency effects on correlation 
functions are presented in a formal mathematical setting that leaves little room for misinterpretation. We also introduce a new weight method that can be used to study correlation functions in the context of experiments where the acceptance and the detection efficiency are complex functions of event and detector characteristics such as the vertex position, the instantaneous beam luminosity, and the detector occupancy. 

This paper is organized as follows. Section \ref{CorrelationDef} discusses the inherent impossibility of distinguishing correlated and uncorrelated particles ab-initio and introduces a generic definition of correlation functions. Section \ref{sec:correlationDecomposition} briefly discusses key properties of
correlation functions and their decomposability in cases where multiple statistically independent sources contribute to particle production. We 
argue that, in practice, such conditions are not realized in elementary particle collisions because conservation laws (energy, momentum, charge, strangeness, etc) implicitly introduce correlation between all produced particles.  Section \ref{instrumentalEffects} presents a discussion of acceptance effects, more
particularly acceptance averaging, and corrections for detection efficiency where we introduce two basic methods (so called Methods 1 and 2)
to determine correlation functions in relative particle pair coordinates such as $\Delta\eta$, the difference of pseudorapidities, or $\Delta\phi$, 
the difference of azimuthal angles. Method 1 is essentially based on a ratio of one dimensional (1D)  histograms of pair production determined as a 
function of $\Delta\eta$ explicitly from particle pairs of same and mixed events. Method 2 is based on ratios of two dimensional (2D) histograms that are functions of the
particle pseudo rapidities ($\eta_1$ and $\eta_2$) separately, and then projected onto $\Delta\eta$. We analyze the merits of the two methods when faced with detection efficiencies that vary through the 
experimental fiducial acceptance in sec. \ref{sec:comparison} and show that Method 2 is by construction robust when the pair 
efficiency factorizes into a product of single particle efficiencies, while Method 1 is simply not robust. We discuss two cases that 
break the factorability of the pair efficiency. In sec. \ref{sec:efficiencyVsZ}, we use a simple model of particle efficiency that depends on the collision vertex
position and show that it breaks the factorability of pair efficiency. The robustness of the correlation function is consequently lost. We however 
present a simple binning technique that enables us to recover the robustness with an arbitrary level of precision in the context of Method 2. We extend 
the discussion of the breaking of the factorability of pair efficiency to other global event observables, such as the detector occupancy and machine luminosity. Last, in sec. \ref{sec:efficiencyVsPhi}, we introduce a general weight method amenable to number and momentum correlation 
functions that simplifies the analysis of correlation functions when global observable such as the event position or detector occupancy 
explicitly break the factorability of pair efficiencies and the robustness of correlation functions. 

\section{Correlated vs Uncorrelated Particles}
\label{CorrelationDef}

The concept of particle correlation is rooted in the notion that if one observes a particle at a given azimuthal angle, rapidity, and transverse momentum, then there is a finite probability to observe one or several other particles at a different azimuthal angle, rapidity, and $p_T$. But the number of particles observed in a "bin" of size $d\vec{p}$ centered at  $\vec{p}=(\phi, \eta, p_{\rm T})$ is intrinsically a stochastic phenomena. 
It is not feasible ab-initio to "tag" particles into a particular "bin" as correlated or uncorrelated. The notion of correlation is a statistical concept
which is meaningful only when all other particles produced in a given collision are taken into account. 
One must resort to statistical techniques to determine the degree of variability and correlation between the yield of particles emitted at two sets of coordinates, or in two regions of phase space. Let $N(\vec{p}_i)$ express the number of particles emitted in a bin "centered" at $\vec{p}_i$. We assume this number fluctuates collision by collision according to some measurable, but likely unknown, probability density function (PDF), $P_1(N(\vec{p}_i))$. The average of this number, $\la N(\vec{p}_i) \ra$, is given by the expectation value of the PDF.
\begin{eqnarray} 
\la N(\vec{p}_i) \ra = E[ N(\vec{p}_i) ] = \int N(\vec{p}_i) P(N(\vec{p}_i)) d(N(\vec{p}_i)) 
\end{eqnarray} 
where we have assumed the normalization $\int P(N) dN = 1$. The variance of the yield is by definition
\begin{eqnarray} 
Var[N] = E[N^2] - (E[N])^2
\end{eqnarray} 
If $m$ distinct and independent production mechanisms are at play, one can demonstrate straightforwardly that the average single particle yield is equal to the sum of the average yield of each of these $m$ production mechanisms or sources. Noting the yield of each source as $N^{(j)}(\vec{p}_i)$, one finds the average yield is simply the sum of the averages. 
\begin{eqnarray} 
\la N(\vec{p}_i)\ra = \sum_{j=1}^{m} \la N^{(j)}(\vec{p}_i) \ra
\end{eqnarray} 
This means the produced yield can in fact be decomposed. But it is crucial to realize that the decomposition does not involve uncorrelated and correlated particles but production from different sources assumed to be independent and uncorrelated. Since each source produces a relatively small number of particles, conservation laws dictate that within each source, particles are likely correlated to some degree. It is also important to realize that the notion of independent, uncorrelated sources, is an artificial construct that does not account for conservation laws as we discuss below.

The number of particles jointly observed at two positions of phase space  $\vec{p}_1$ and $\vec{p}_2$ may likewise be characterized by some joint probability density function, $P_2(N(\vec{p}_1), N(\vec{p}_2))$, assumed to be measurable but a priori  unknown. Statistically, the number of particles produced at $\vec{p}_1$ and $\vec{p}_2$ can be considered uncorrelated, or statistically independent, if and only if the joint probability can be factorized. 
\begin{eqnarray} 
\label{statisticalIndependence}
P_{2,uncor}(N(\vec{p}_1), N(\vec{p}_2))
= P_{1}(N(\vec{p}_1)) \times P_{1}(N(\vec{p}_2))
\end{eqnarray} 
A measure of the degree of correlation between produced particles is thus achieved by comparing $P_2(N(\vec{p}_1), N(\vec{p}_2))$ with $P_{2,uncor}(N(\vec{p}_1), N(\vec{p}_2)) $.
Indeed, the difference between these two probability densities expresses in full details the likelihood of correlation between the two emission directions. The difference is null if the emission taking place at the two directions are totally uncorrelated and non null if some form of correlation exists. 
Though it is possible to regard a collision as a superposition of several disjoint processes, it is not a priori possible to know the degree to which particles of a given process are correlated or uncorrelated unless  perhaps one can specifically model the particle production  in terms of a succession of specific elementary processes, which uniquely specify $P_2(N(\vec{p}_1), N(\vec{p}_2))$. Measuring $P_{2,corr} $ for all values of $N(\vec{p}_1) $  and $N(\vec{p}_2) $ at all relevant phase space positions is a rather arduous task rarely considered. It is however possible to estimate whether the yields at $N(\vec{p}_1) $  and $N(\vec{p}_2) $ are correlated by measuring specific moments of $P_2(N(\vec{p}_1),N(\vec{p}_2)) $. The most basic of these moments is the covariance of $N(\vec{p}_1) $  and $N(\vec{p}_2) $. We introduce the convenient shorthand notation $N(i) = N(\vec{p}_i) $. The covariance can be written:
\begin{eqnarray} 
\label{covDef}
Cov[N(1),N(2)] = E[ N(1)N(2)] - E[ N(1)]  E[N(2)]
\end{eqnarray} 
By definition, the covariance is null if the yields $N(\vec{p}_1)$ and $N(\vec{p}_2)$ are statistically independent. Deviation from zero implies the two quantities are correlated (or anti-correlated). Note however that a null covariance is not a sufficient condition to conclude the quantities are statistically independent. The proper condition is factorability of $P_2$ into $P_1\times P_1$. 
It is convenient to introduce the following shorthand notations:
\begin{widetext}
\begin{eqnarray} 
\rho_1(\vec{p}_i) &=& \la N(i) \ra = \int N(i) P_1(N(i)) dN(i)\\ \nonumber
\rho_2(\vec{p}_i,\vec{p}_j) &=& \la N(i) N(j) \ra = \int N(i) N(j) P_2(N(i),N(j)) dN(i)dN(j)
\end{eqnarray} 
\end{widetext}
The quantity $\rho_1(\vec{p}_i)$ corresponds to the average of the number of particles, $N(\vec{p}_1)$, produced in a finite size bin centered at $\vec{p}_i$. Note that this number can be obtained by integration over a certain portion of the phase space. For instance, in the context of $\Delta \eta, \Delta \phi$ correlation studies, one may integrate over a finite $p_{\rm T}$  range while using finite size bins $\delta\eta$ and $\delta\phi$ in pseudo rapidity and azimuth respectively.
\begin{eqnarray} 
\rho_1(\eta_i, \phi_i) = \int_{\delta\phi} {\int_{\delta\eta} {\int_{p_{T,min}}^{p_{T,max}}  {\frac{d^3N}{p_{\rm T} d\phi d\eta dp_{\rm T}} d\phi d\eta dp_{\rm T}}}}
\end{eqnarray} 
Similarly, the quantity $\rho_2(\vec{p}_i,\vec{p}_j)$ corresponds to the average of the number of pairs $N(\vec{p}_i) N(\vec{p}_j)$ detected jointly (i.e. event by event) in two finite size bins. 
\begin{eqnarray} 
\rho_2(\eta_1, \phi_1,\eta_2, \phi_2) &=& \int_{\delta\phi_1} {\int_{\delta\eta_1} {\int_{p_{T,1,min}}^{p_{T,1,max}}  }}\\ \nonumber
 & &\int_{\delta\phi_2} {\int_{\delta\eta_2} {\int_{p_{T,2,min}}^{p_{T,2,max}} }} \\ \nonumber
 & &{\frac{d^6N}{dp_1^3 dp_2^3} d\phi_1 d\eta_1 dp_{{\rm T},1}d\phi_2 d\eta_2 dp_{{\rm T},2} }
\end{eqnarray} 
The covariance (Eq. \ref{covDef}) may then be written in a familiar and generic form:
\begin{eqnarray} 
{\rm Cov}[N(\vec{p}_1),N(\vec{p}_2)] = \rho_2(\vec{p}_1,\vec{p}_2) - \rho_1(\vec{p}_1)\times \rho_1(\vec{p}_2)
\end{eqnarray} 
Since the fluctuating particle yields $N(\vec{p}_1)$ and $N(\vec{p}_2)$ are functions of the coordinates (bins), the covariance is really a function of these coordinates.  We thus define the generic correlation function of two particles as of function of their momenta
$\vec{p}_1$ and $\vec{p}_2$ as:
\begin{eqnarray} 
C_2(\vec{p}_1,\vec{p}_2) = \rho_2(\vec{p}_1,\vec{p}_2) - \rho_1(\vec{p}_1)\times \rho_1(\vec{p}_2)
\end{eqnarray} 
In the context of a $\Delta \eta$ vs. $ \Delta \phi$ type analysis, this function can be written:
\begin{eqnarray} 
C_2(\eta_1, \phi_1,\eta_2, \phi_2) &=& \rho_2(\eta_1, \phi_1,\eta_2, \phi_2) \\ \nonumber
 & & - \rho_1(\eta_1, \phi_1)\times \rho_1(\eta_2, \phi_2) 
\end{eqnarray} 
where it is understood that single particle and pair yields are measured in finite size bins centered at $\eta_1, \phi_1$ and $\eta_2, \phi_2$. This {\it correlation function} is the basis for much of the recent analyses carried out in heavy ion physics at RHIC and LHC~\cite{Agakishiev2013a,Abelev2009a,Adams2007a,Abelev2013a,Aad2013a,Chatrchyan2013b}. 

Both the strength and shape of this type of correlation function carry important information about the reaction mechanism of particle production. One approach sometimes used to interpret the strength of a covariance is to compare it to the square root of the product of the variances of the two variables. This can be accomplished by considering the Pearson coefficient, $p(\vec{p}_1, \vec{p}_2)$, defined as 
\begin{eqnarray} 
p(\vec{p}_1, \vec{p}_2) = \frac{ {\rm Cov}[N(\vec{p}_1),N(\vec{p}_2)] }{{\rm Var}[N(\vec{p}_1)]^{1/2}{\rm Var}[N(\vec{p}_2)]^{1/2}}
\end{eqnarray} 
By construction, this ratio is bound to the interval $[-1,1]$ and provides, in principle, a convenient way to gauge the strength of the correlation between the particle emission at $\vec{p}_1$ and $\vec{p}_2$. In practice, one should note that instrumental effects  may render the evaluation of ${\rm Var}[N(\vec{p}_i)]$ rather challenging. Indeed, the variance of the yield depends both on the physics and the efficiency of detection in a non trivial way. Note that the practice of substituting $\sqrt{N(\vec{p}_i)}$ for ${\rm Var}[N(\vec{p}_1)]^{1/2}$ is based on the assumption that the process is Poissonian and such that the variance is equal to the mean. This assumption is however {\it invalid} because of two reasons. The first reason is that  particle emission is in general correlated at some level and is consequently not perfectly poissonian. The second  is that finite efficiency implies binomial sampling whose variance is not equal to the mean. 
A second approach consists in simply comparing the covariance with the product of the single particle yields. One thus defines the normalized correlation function $R_2$ as follows:
\begin{eqnarray} 
\label{eq:rdef}
R_2(\vec{p}_1, \vec{p}_2) &=& \frac{ {\rm Cov}[N(\vec{p}_1),N(\vec{p}_2)] }{\la N(\vec{p}_1)\ra \la N(\vec{p}_2)\ra} \\ \nonumber
&=&  \frac{\rho_2(\vec{p}_1,\vec{p}_2) - \rho_1(\vec{p}_1)\rho_1(\vec{p}_2)}{ \rho_1(\vec{p}_1)\rho_1(\vec{p}_2)} \\ \nonumber
&=& \frac{\rho_2(\vec{p}_1,\vec{p}_2)}{\rho_1(\vec{p}_1)\rho_1(\vec{p}_2)} - 1
\end{eqnarray} 
This definition involves several practical advantages and is commonly used in mixed event analyses to account for instrumental effects. We discuss the decomposition of this correlation function as a sum of the correlation functions of a superposition of multiple processes in sec. \ref{sec:correlationDecomposition}. We discuss acceptance and efficiency effects on this correlation in sec. \ref{sec:acceptance} and sec. \ref{sec:efficiency} respectively. 

\section{Correlation function decomposition}
\label{sec:correlationDecomposition}

In order to extract a physical interpretation, several (published) analyses have adopted a two component decomposition of the function $R_2$. It is often assumed that the lowest point of two particle densities corresponds to an uncorrelated yield. Many authors  thus subtract  the yield at the minimum as if it corresponded to a zero correlation yield. The remainder is then considered as the true correlated signal, analyzed (fitted), and interpreted as such.This procedure is unfortunately rather artificial and may lead to  misunderstanding of the measured correlations. Strictly speaking, the correlation $R_2$ is defined on the basis of the notion of covariance. To seek a level where there are no correlations, i.e. statistical independence, thus requires the covariance to be properly normalized and null. However, a null covariance is not a sufficient condition for statistical independence, there is thus no satisfactory way to subtract a background and separate correlated and uncorrelated particle strengths. This is simply not feasible in a consistent way. Proponents of the method might however argue that one can legitimately decompose the correlation function $R_2$ into several components provided particle production proceeds via a superposition of independent processes. To accomplish this, assume particle production can indeed be described as a superposition of $m$ distinct and disjoint (i.e. statistically independent) processes. To simplify, let us assume the processes are identical and have single particle and particle pair yields noted $\rho_1^{1}(\vec{p}_i)$ and $\rho_2^{1}(\vec{p}_1,\vec{p}_2)$ respectively. It is  straightforward to show that the superposition of $m$ identical and statistically independent such processes leads to the following single particle and pair yields. 
\begin{eqnarray} 
\label{eq:decomposition}
\rho_1^{(m)}(\vec{p}_i) &=& \sum_{i=1}^{m} \rho_1^{1}(\vec{p}_i) =  m \rho_1^{1}(\vec{p}_i)\\ \nonumber
\rho_2^{(m)}(\vec{p}_1,\vec{p}_2) &=& \sum_{i=1}^{m} \rho_2^{1}(\vec{p}_1,\vec{p}_2)  + \sum_{i\neq j=1}^{m} \rho_1^{1}(\vec{p}_1) \rho_1^{1}(\vec{p}_2) \\ \nonumber
&=& m \rho_2^{1}(\vec{p}_1,\vec{p}_2) + m(m-1) \rho_1^{1}(\vec{p}_1) \rho_1^{1}(\vec{p}_2)
\end{eqnarray}  
The above expression for $\rho_2^{m}(\vec{p}_1,\vec{p}_2)$ obviously provides ground for a two component separation of the correlation strength. Indeed, the first term represents the correlated part of the signal and the second term is a combinatoric part formed from pairs of particles produced by distinct processes (assumed to be uncorrelated). The use of this expression in analyses of $A+A$ collision data is however legitimate only if the pair yield produced in, say, central collisions is in fact a superposition of $m$ simpler $p+p$ processes. But correlations measured in $A+A$ collisions have manifestly a different shape than those observed in $p+p$. The measured correlations are thus obviously 
not a superposition of unmodified and independent $p+p$ (or even parton-parton) processes. The observed correlations 
may in principle have collective origins (e.g. hydrodynamic flow), involve a superposition of modified correlation functions (from $p+p$ or
parton-parton collisions), or both. This particular decomposition is therefore  not strictly applicable.  

In a finite collision system, be it $p+p$, $Au+Au$, or $Pb+Pb$, laws of conservations readily imply that the production of all particles are correlated at some level. The conservation of energy and momentum imply in particular that if one or several particles are found in a given momentum bin, the momenta of other particles 
are intrinsically constrained at some level. This is readily obvious for a resonant particle undergoing a two-body decay at rest in the lab frame. The detection of one particle immediately tells us where the second is going. In a three prong decay, the correlation is weaker but nonetheless present. Energy momentum conservation constrains the position of the other two particles. The same conclusion applies for a system of 10, 100, or several 1000 produced particles. The constraints are weaker for an increasing large number of particles but they never vanish \cite{Borghini2002a,Borghini2007a}. Other conservation laws also generate intrinsic correlations of the produced particles. For instance, charge conservation implies the production of a positive particle must be  accompanied by the production of a negative particle. This is true whether 10, 100, or 1000 pairs of positive and negative particles are produced. The fact that particles are emitted over an extended range of pseudorapidities and azimuth only dilutes the point-to-point strength of the correlation, it does not eliminate it. Conservation of other quantum numbers produce correlations as well. In essence, all produced particles are effectively correlated to some other particles, it is therefore strictly not logically consistent to attempt a decomposition of $R_2$ (or the single particle yield) into correlated and uncorrelated parts.

\section{Accounting for instrumental effects}
\label{instrumentalEffects}

Measurements of two-particle densities and correlation function are affected by  various experimental conditions, and must in principle be explicitly corrected for instrumental effects such as detector acceptance, resolution, detection efficiency, contamination, etc. We first discuss 
the effects of the acceptance and acceptance averaging in Sec.~\ref{sec:acceptance}. Effects associated with detection efficiencies
are discussed in following sections. A discussion of resolution, contamination and other instrumental effects is outside the scope of this paper.

\subsection{Acceptance Averaging }
\label{sec:acceptance}

The acceptance used in measurements of correlation functions obviously impacts the outcome of measurements. The effects of the acceptance should however not be mistaken with those associated with detection efficiency. Effects associated with detection efficiency can be eliminated or at the very least suppressed using techniques discussed in the next section. Acceptance effects cannot be corrected for as easily. One can at best make educated extrapolations. In the context of measurements of single particle spectra (e.g. vs $p_{\rm T}$ ), such extrapolations are usually meant to extend a measurement down to zero transverse momentum. Extrapolating a measurement outside of the kinematical region where it is taken amounts to "inventing" information. But physically motivated parametric models of the data can often be used to extend a measurement down to zero $p_{\rm T}$ . Estimation of systematic uncertainties then essentially amount to "guessing" the range of models and parameter values that are consistent with the data actually measured. 

Correcting for acceptance effects in measurements of a correlation function is far more subtle and is usually not attempted. A procedure commonly done however is to average over unmeasured parameters. This is the case of two particle correlation  functions measured as function of the relative pseudo rapidity of the particles, and optionally as a function of their relative azimuth of emission. In such analyses, the individual rapidities, $\eta_1$ and $\eta_2$, are combined to determine the relative or difference in pseudo rapidity $\Delta \eta = \eta_1 - \eta_2$.  In general, however, there is no reason to assume the correlation function depends only 
on the difference, $\Delta \eta$, it might also depend on the actual values of $\eta_1$ and $\eta_2$, or their mean. Two-particle correlation functions should really be studied explicitly in terms of $\eta_1$ and $\eta_2$ or, alternatively, as a function of the pseudo rapidity difference, $\Delta \eta=\eta_1 - \eta_2$, and the average pseudo rapidity, $\overline{\eta}=(\eta_1 +\eta_2)/2$.  If the average pseudo rapidity, $\overline{\eta}=(\eta_1 +\eta_2)/2$ dependence is weak or of limited interest, it may then be averaged out.

A measurement of the two-particle density $\rho_2( \eta_1,\eta_2) $ can be recast as a measurement of $\rho_2( \Delta \eta,\overline{\eta}) $ for which the Jacobian $|\partial(\Delta\eta,\overline{\eta})/\partial(\eta_1,\eta_2)|$ is equal to unity. It is thus legitimate to consider measurements of correlation functions $C_2(\Delta \eta,\overline{\eta})$ and $R_2(\Delta \eta,\overline{\eta})$ or restrict these measurements to their $\Delta \eta$ dependence only by averaging over $\overline{\eta}$. To simplify, let us assume that both particles are measured in the same pseudo rapidity range $-\eta_o \le \eta < \eta_o$. Averaging over $\overline{\eta}$ is then calculated as follows:
\begin{eqnarray} 
C_2(\Delta \eta) = \frac{1}{\Omega(\Delta)} \int_{-(\eta_o -\Delta\eta/2)}^{\eta_o -\Delta\eta/2} C_2(\Delta \eta,\overline{\eta}) d\overline{\eta}
\end{eqnarray} 
where $\Omega(\Delta\eta) = 2\eta_o - \Delta \eta$ is an acceptance factor accounting for the fact there are far many more ways to measure $\Delta\eta=0$ than to measure $\Delta \eta=2\eta_o$.  This type of measurement has routinely been carried out in recent years. Unfortunately, since the averaging over $\overline{\eta}$ is often combined with correction for detection efficiency, some confusion may and has in fact arisen in some recent works. This point is discussed in sec. \ref{sec:efficiency2} after techniques for correction efficiency are discussed in the next section. 

\subsection{Efficiency and Observable Robustness}
\label{sec:efficiency}

Let  $n(\vec{p_i})$ and $N(\vec{p_i})$ respectively represent the number of particle measured and actually produced by the collision system at the given momentum (or in a "bin" centered at that momentum). We will here neglect resolution effects on the measurement of $\vec{p_i}$. The number of detected particles is typically smaller than the number of produced particles owing to particle loses in the detection and reconstruction of the events. For a well behaved detector, the likelihood of detecting (or not detecting) a particle can be described by an efficiency. The detection process is equivalent to a sampling where the probability of keeping a given particle is determined by the detection efficiency, $\epsilon$. The probability of observing "n" particles given "N" were produced, noted $P_{det}(n|N;\epsilon)$,  can then be described by a binomial distribution. 
\begin{eqnarray} 
P_{det}(n|N;\epsilon) = \frac{\epsilon^n (1-\epsilon)^{N-n}}{n! (N-n)!}
\end{eqnarray} 
Let $P_m(n)$ express the probability of measuring $n$ particles  and $P_p(N)$ the probability of producing N particles at $\vec{p_i}$. The probability $P_m$ may readily be expressed in terms of $P_p$ and $P_{det}$ as follows:
\begin{eqnarray} 
\label{pm}
P_m(n) = \int P_{det}(n|N;\epsilon) P_p(N) dN
\end{eqnarray}  
The mean number of produced and detected particles are calculated from their respective probability distributions.
\begin{eqnarray} 
\label{avgs}
\la N \ra = \int N P_p(N) dN \\ \nonumber
\la n \ra = \int n P_m(n) dn 
\end{eqnarray} 
A simple relationship obviously exists between the two averages. Substitute the expression for $P_m$ in the calculation of the mean $\la n \ra$.
\begin{eqnarray} 
\la n \ra = \int ndn  \int P_{det}(n|N,\epsilon) P_p(N) dN 
\end{eqnarray} 
The order of integration does not matter. One then obtains the well known result:
\begin{eqnarray} 
\la n \ra &=& \int dN P_p(N) \int n P_{det}(n|N,\epsilon)  dn \\ \nonumber
&=&  \epsilon \int  NdN P_p(N) \\ \nonumber
&=& \epsilon \la N\ra
\end{eqnarray} 

The same exact analysis can be repeated for measurements of pairs. Let $P_m(n(1),n(2))$ express the probability of measuring $n(1)$ and $n(2)$ particles 
at $\vec{p_1}$ and $\vec{p_2}$ respectively. The corresponding probability for the number of produced particles $N(1)$ and $N(2)$ is noted $P_p(N(1),N(2))$.
The probability to simultaneously detect $n(1)$ and $n(2)$  particles when $N(1)$ and $N(2)$ are produced is described with some joint detection probability noted $P_{det}(n(1),n(2)|N(1),N(2))$. We thus repeat the analysis carried for single particles. The mean number of produced and measured number of pairs are as follows:
\begin{widetext}
\begin{eqnarray} 
\la N(1) N(2) \ra = \int N(1) N(2) P_p(N(1),N(2)) dN(1) dN(2) \\ \nonumber
\la n(1) n(2) \ra  = \int  n(1) n(2)  P_m(n(1),n(2) ) dn(1) dn(2)  
\end{eqnarray} 
The measured and produced number of pairs are related according to
\begin{eqnarray} 
\la n(1) n(2) \ra =  \int  n(1) n(2) dn(1) dn(2) \int  P_{det}(n(1),n(2)|N(1),N(2)) P_p(N(1),N(2)) dN(1)dN(2)
\end{eqnarray}  
Clearly, if the probability of simultaneously detecting particles at $\vec{p_1}$ and $\vec{p_2}$ can be factorized, i.e. for
\begin{eqnarray} 
P_{ideal-det}(n(1),n(2)|N(1),N(2)) = P_{det}(n(1)|N(1),\epsilon(1)) P_{det}(n(2)|N(2),\epsilon(2))
\end{eqnarray} 
\end{widetext}
where we assume the detection efficiencies at $\vec{p_1}$ and $\vec{p_2}$ may differ. We then get
\begin{eqnarray} 
\la n(1) n(2) \ra =  \epsilon(1) \epsilon(2) \la N(1)N(2)\ra
\end{eqnarray} 
In this context, one finds that the measured correlation function $R_2$ is independent of detection efficiencies. The ratio of measured pair to product of measured singles equals the ratio of produced pairs to the product of produced singles. \begin{eqnarray} 
\label{r2robust}
R_2(\vec{p_1},\vec{p_2}) &=& \frac{\la n(1) n(2)\ra}{ \la n_1\ra \la n(2)\ra}- 1 \\ \nonumber
&=& \frac{\epsilon(1) \epsilon(2) \la N(1) N(2)\ra}{\epsilon(1) \la N(1) \ra\epsilon(2)\la N(2)\ra} - 1\\ \nonumber
&=& \frac{\la N(1) N(2)\ra}{\la N(1) \ra \la N(2)\ra}-1
\end{eqnarray} 
The observable $R_2$ is therefore said to be "robust". By contrast, the correlation function $C_2(\vec{p_1},\vec{p_2})$ explicitly 
depends on the efficiencies and is consequently not robust. It must be corrected for particle losses associated with the efficiencies $\epsilon(1)$ and $ \epsilon(2)$   which are in general functions of the measured momentum coordinates $\vec{p_1}$ and
$\vec{p_2}$. This may be accomplished by invoking the robust correlation function $R_2$ explicitly or by carrying out mixed event analysis. The pair yield obtained in mixed events is equivalent to the denominator of Eq. \ref{r2robust} since the particle yields 
detected are not correlated by a common production mechanism and only depend on the product of the efficiencies $ \epsilon(1) \epsilon(2)$. Mixed event analyses thus provide, in effect, an estimate of $R_2$ which can then be normalized to yield $C_2$. It is however easy to lose track of the distinct effects of acceptance and efficiency. Improper averaging of the correlation on $\overline{\eta}$ may in particular yield pathological results. This is discussed in the following section.

\subsection{Efficiency Correction and Acceptance}
\label{sec:efficiency2}

Eq. \ref{r2robust} provides a technique to obtain an efficiency corrected correlation function but it is a function of coordinates $\vec{p_1}$
and $\vec{p_2}$. How then does one obtain correlation functions that depends, say, only on  the difference of the pseudo rapidity of the two particles, $\Delta \eta$. One technique commonly used is to carry the analysis of real and mixing events using 1D histograms. Let $H_1^{(real)}(\Delta \eta)$ and $H_1^{(mixed)}(\Delta \eta)$ respectively represent the pair yield as a function of $\Delta \eta$ with some suitable binning. The correlation function $R_2(\Delta \eta)$ may then be estimated by taking the ratio of the "real" and "mixed" pairs as follows.
\begin{eqnarray} 
R_2(\Delta \eta) = \frac{H_1^{(real)}(\Delta \eta)}{H_1^{(mixed)}(\Delta \eta)}
\end{eqnarray} 
We shall refer to this technique as Method 1 in the following. As we demonstrate next, one can expect this expression to yield a reasonable estimate of the actual correlation function provided the efficiencies $\epsilon(1)$ and $ \epsilon(2)$ are not rapidly changing with the coordinates $\eta_1$ and $\eta_2$. The estimate may however be of limited precision if the efficiency varies considerably over the acceptance of the measurement as is often the case in situations where the detector occupancy varies with $\eta$ or wherever "edge effects" are important. 

An alternative technique, which we refer to as Method 2 in the following, involves a measurement of the correlation  as a function of  $\eta_1$ and $\eta_2$ explicitly, $R_2(\eta_1,\eta_2)$, using 2D histograms $H_2^{(real)}(\eta_1,\eta_2)$ and $H_2^{(mixed)}(\eta_1,\eta_2)$ for real and mixed pairs respectively.
\begin{eqnarray} 
R_2(\eta_1,\eta_2) = \frac{H_2^{(real)}(\eta_1,\eta_2)}{H_2^{(mixed)}(\eta_1,\eta_2)}
\end{eqnarray} 
 The histograms $H_2^{(real)}(\eta_1,\eta_2)$ and $H_2^{(mixed)}(\eta_1,\eta_2)$, when properly normalized by the number of events, represent respectively the number of particle pairs $ \la n(1) n(2)\ra$ and product of singles $ \la n(1)\ra \la n(2)\ra$ measured at $\eta_1$ and $\eta_2$. These numbers are subject to
 efficiencies, as discussed above. If the pair efficiency factorizes, then $R_2(\eta_1,\eta_2)$ is perfectly robust, i.e. its definition yields a perfect efficiency corrected result. The $\Delta \eta$ dependence of the correlation function is then obtained by a change of variables and averaging over   $\overline{\eta}$.
\begin{widetext}
\begin{eqnarray} 
R_2(\Delta \eta) = \frac{1}{\Omega(\Delta \eta)} \int R_2(\eta_1,\eta_2) \delta(\Delta\eta -\eta_1+\eta_2) d\eta_1 d\eta_2
\end{eqnarray} 
\end{widetext}
where $\Omega(\Delta \eta)$ accounts for the size of the acceptance for $\Delta \eta$.
Both method 1 and 2  yield  a correlation function averaged over $\overline{\eta}$. By construction, the correlation function obtained with  Method 2 is perfectly robust whenever the efficiency factorizes. It may however suffers from a weak geometrical effect, known as aliasing, if the 
binning used is too coarse. We will show that Method1 is not strictly robust although it may yield reasonably accurate results if the detection efficiency is constant or varies weakly throughout the detector acceptance. 

\section{Comparison of Methods 1 \& 2}
\label{sec:comparison}

\begin{figure}[t!]
\centering
\includegraphics[width=0.45\textwidth]{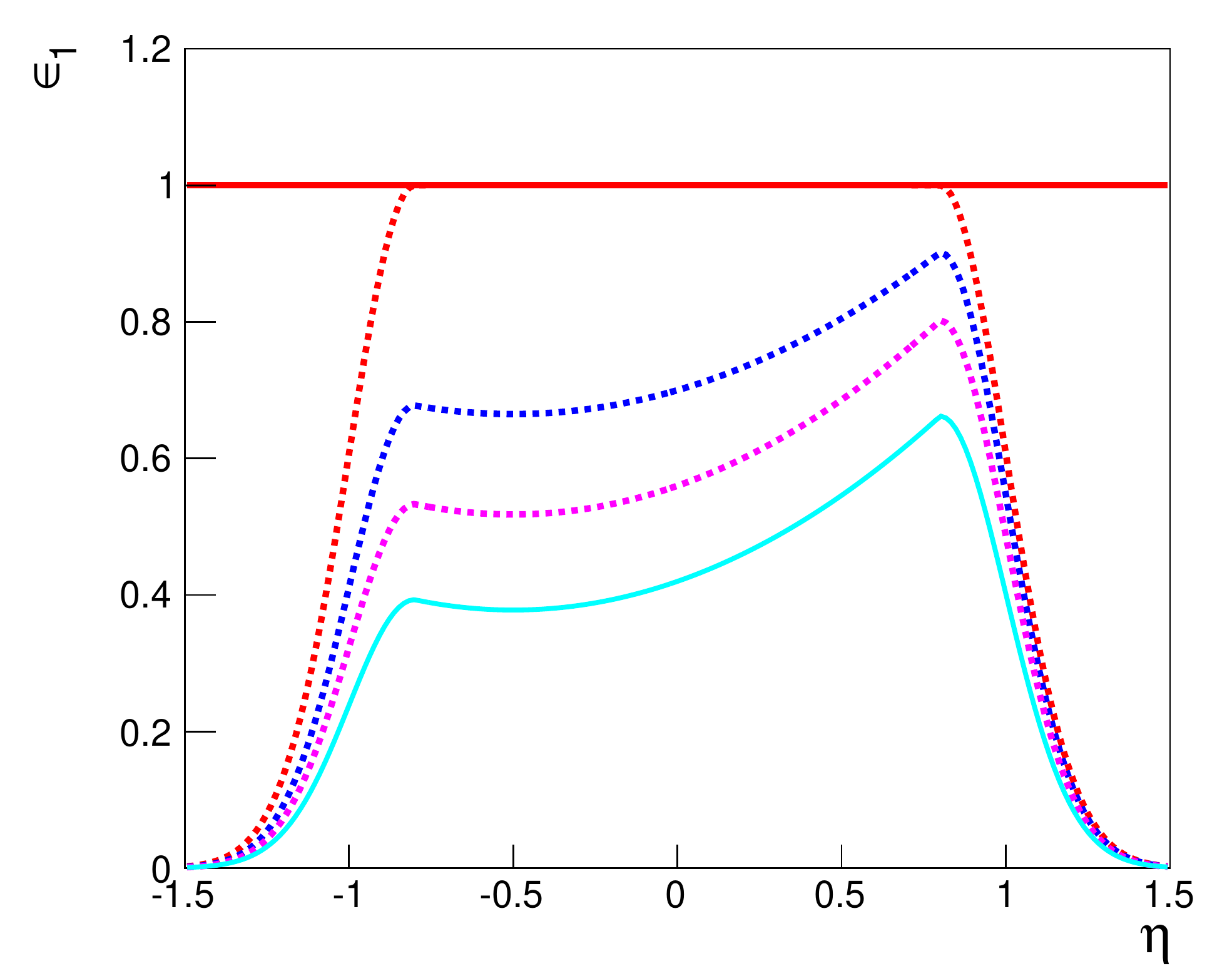}
\caption{(Color Online) Modeling of the detection efficiency dependence on pseudo rapidity used in sec. \ref{sec:efficiencyVsEta}: perfect efficiency (solid red), flat response with smooth edges (dash red), and non-linear response with edge effects (other curves) as described in the text. }
\label{fig:BasicEfficiencyVsEta}
\end{figure}

Ideally, particle detection efficiency  should be uniform across the fiducial acceptance of a detector. In practice, the efficiency  is  usually a complicated function of particle types (species), their pseudo rapidity, azimuth, and transverse momentum. In collider detectors, the nominal efficiency is usually reasonably large and approximately uniform in a nominal range $-\eta_{o}  \le \eta < \eta_{o} $ beyond which it quickly vanishes.
We model the nominal efficiency according to the following simple model:
\begin{eqnarray} 
\epsilon(\eta) &=&  \epsilon_o e^{-(\eta-\eta_{<})^2/2\sigma_{\epsilon}^2} \hspace{0.15in} \rm{ for\ }  \eta <\eta_{<}  \\ \nonumber
                       &=&  \epsilon_o \hspace{0.981in} \rm{ for\  }  \eta_{<} < \eta < \eta_{>}  \\ \nonumber
                       &=&  \epsilon_o e^{-(\eta-\eta_{>})^2/2\sigma_{\epsilon}^2} \hspace{0.15in} \rm{ for\ }  \eta >\eta_{>}  
\end{eqnarray} 
where  $\epsilon_o$ is the nominal detection efficiency in the fiducial volume while $\eta_{<}=-\eta_o $ and $\eta_{>}=\eta_o $  represent the nominal low and high "edges" of the detector acceptance beyond which the efficiency rolls off and vanishes. The width $\sigma_{\epsilon}$ determines the roll off rate of the efficiency with $\eta$. The model is illustrated in Fig. \ref{fig:BasicEfficiencyVsEta} for $\epsilon = 0.7$, $\eta_{<} =0.8 $ and $\eta_{>} 0.8$ with a red dash curve.
 The nominal efficiency is here defined as the probability to detect particles (as a function of $\eta$) when the collision takes place in the geometric center of the detector. Various factors may however influences or adversely affect the performance of a detection system and 
 consequently modify this response curve. We consider some of these effects and their impact on correlation function in the following subsections. 

\subsection{Efficiency Dependence on Pseudorapidity.}
\label{sec:efficiencyVsEta}

Advanced designs and attention to details during construction provide for reasonably uniform efficiency response within the experimental fiducial volume of modern detectors. The performance is rarely perfect however and small to medium dependencies on pseudo rapidity and azimuthal angles may remain or be introduced by various equipment failures. These can have a significant impact on correlation functions if not properly accounted for. In this section, we consider the effects  of detector efficiency with finite dependence on pseudo rapidity. We modify the nominal efficiency curve (red dash)  shown in Fig. \ref{fig:BasicEfficiencyVsEta} to have an arbitrary quadratic dependence on pseudo rapidity in the range $-\eta_{<} \le \eta < \eta_{>}$ as follows.
\begin{eqnarray} 
\epsilon(\eta) &=&  \epsilon_q(\eta) e^{-(\eta-\eta_{<})^2/2\sigma_{\epsilon}^2} \hspace{0.15in} \rm{ for\ } \eta <\eta_{<}  \\ \nonumber
                       &=&  \epsilon_q(\eta) \hspace{0.981in} \rm{for\ } \eta_{<} < \eta < \eta_{>}  \\ \nonumber
                       &=&  \epsilon_q(\eta) e^{-(\eta-\eta_{>})^2/2\sigma_{\epsilon}^2} \hspace{0.15in} \rm{ for\ }  \eta >\eta_{>}  
\label{eq:epsilonVsEta}
\end{eqnarray} 
where
\begin{eqnarray} 
\epsilon_q (\eta) = 1 + \alpha (\eta - \eta_o) + \beta (\eta - \eta_o)^2
\end{eqnarray} 
The coefficients $\alpha$ and $\beta$ determine the linear and quadratic dependence of the model while $\eta_o$ represents the "center" of the acceptance, taken here to be the origin ($\eta_o=0$). The quadratic model is illustrated in Fig. \ref{fig:BasicEfficiencyVsEta} 
for $\epsilon_o=0.7, \alpha=0.2$, $\beta=0.2$ (blue dash line), $\epsilon_o=0.56, \alpha=0.3$, $\beta=0.3$ (purple dash line),
$\epsilon_o=0.7, \alpha=0.4$, $\beta=0.4$ (light blue solid line). The pair efficiency is here assumed to factorize. Fig. \ref{fig:BasicPairEfficiencyVsEta} 
display the pair efficiency dependence for (a) perfect efficiency, 
(b) a flat response with smooth edges, and non-linear response with edge effects shown for $\epsilon_o=0.7, \alpha=0.4$, $\beta=0.4$.

\begin{widetext}

\begin{figure}[t!]
\centering
\includegraphics[width=0.32\textwidth]{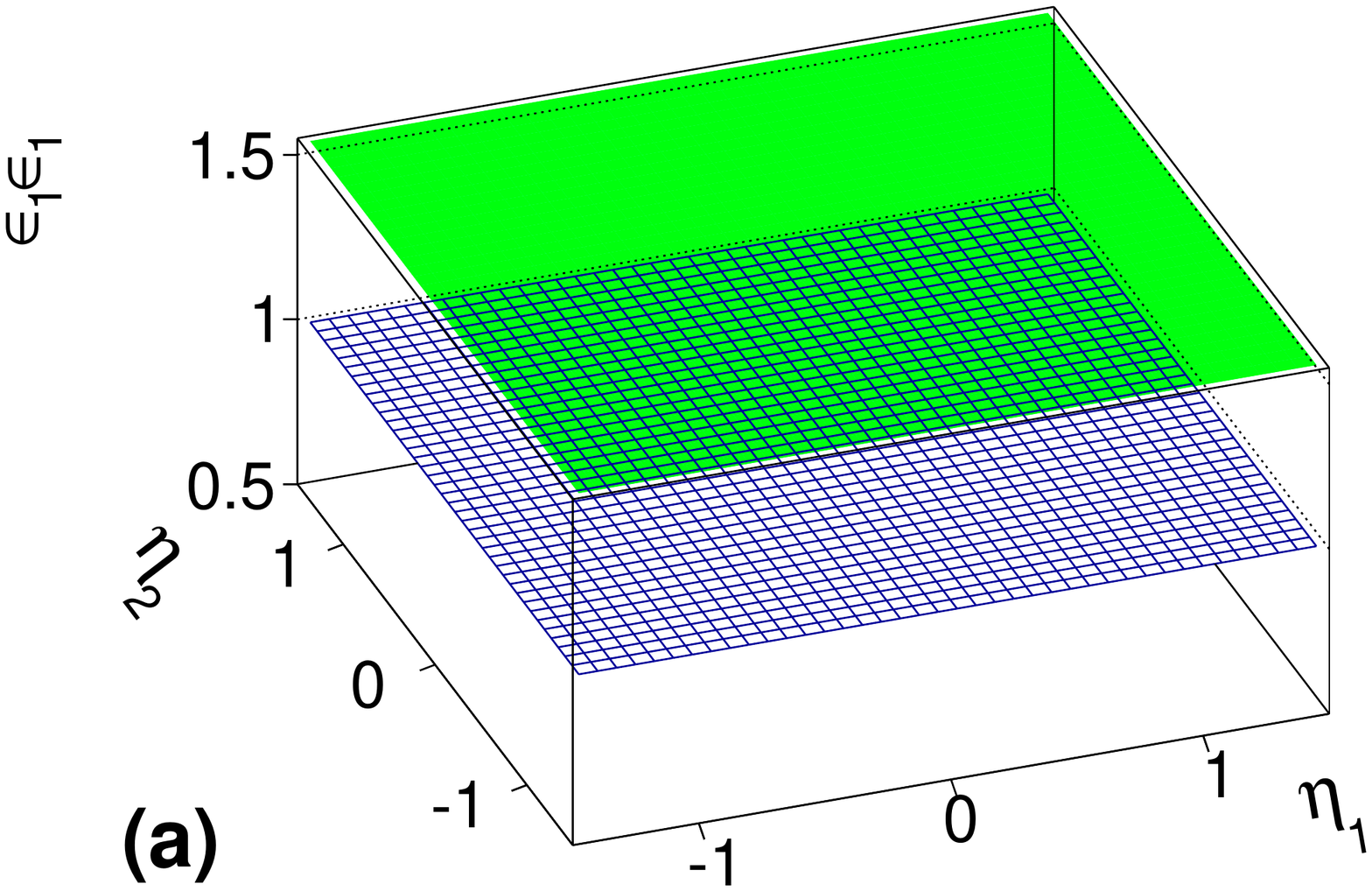}
\includegraphics[width=0.32\textwidth]{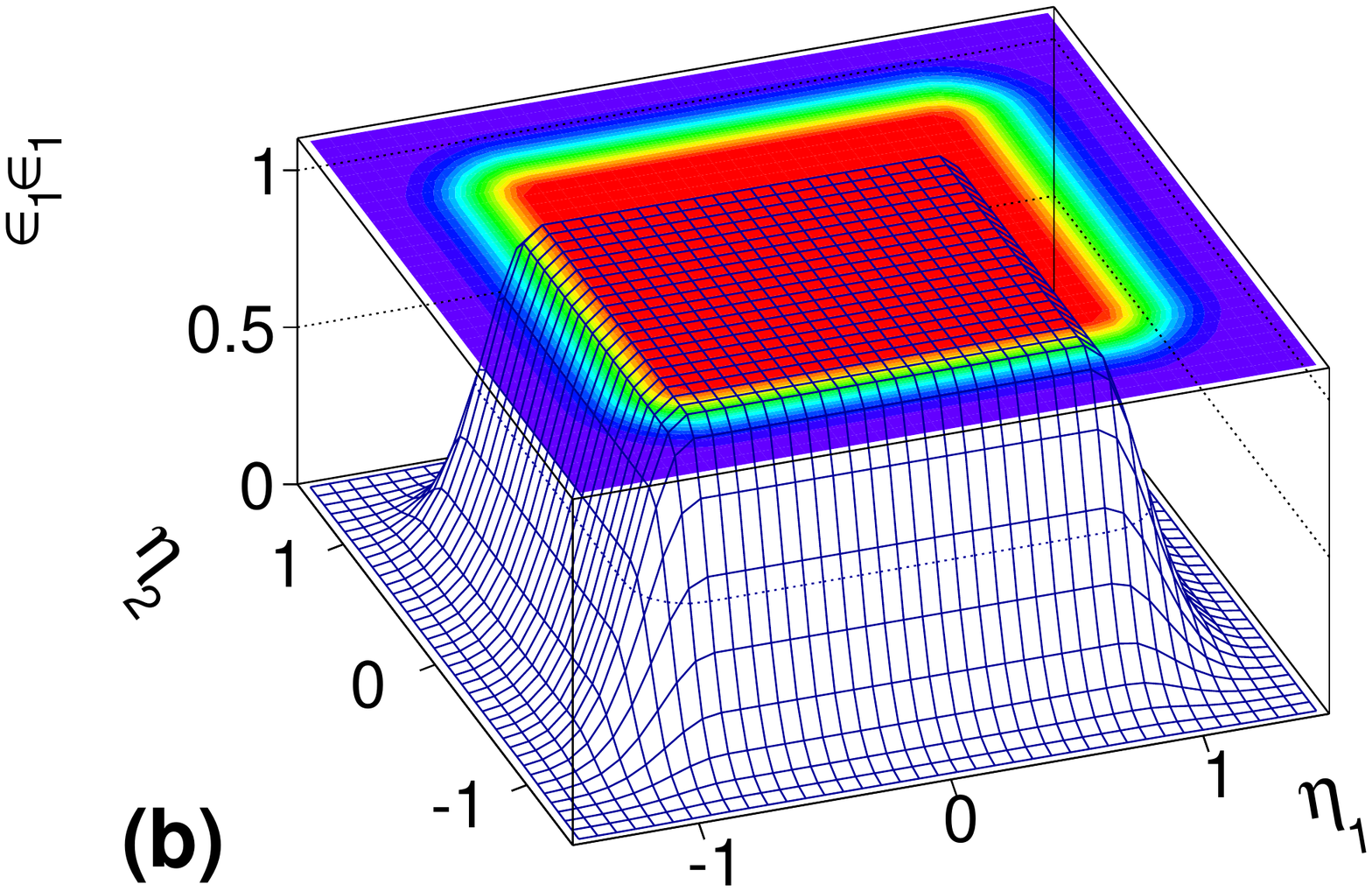}
\includegraphics[width=0.32\textwidth]{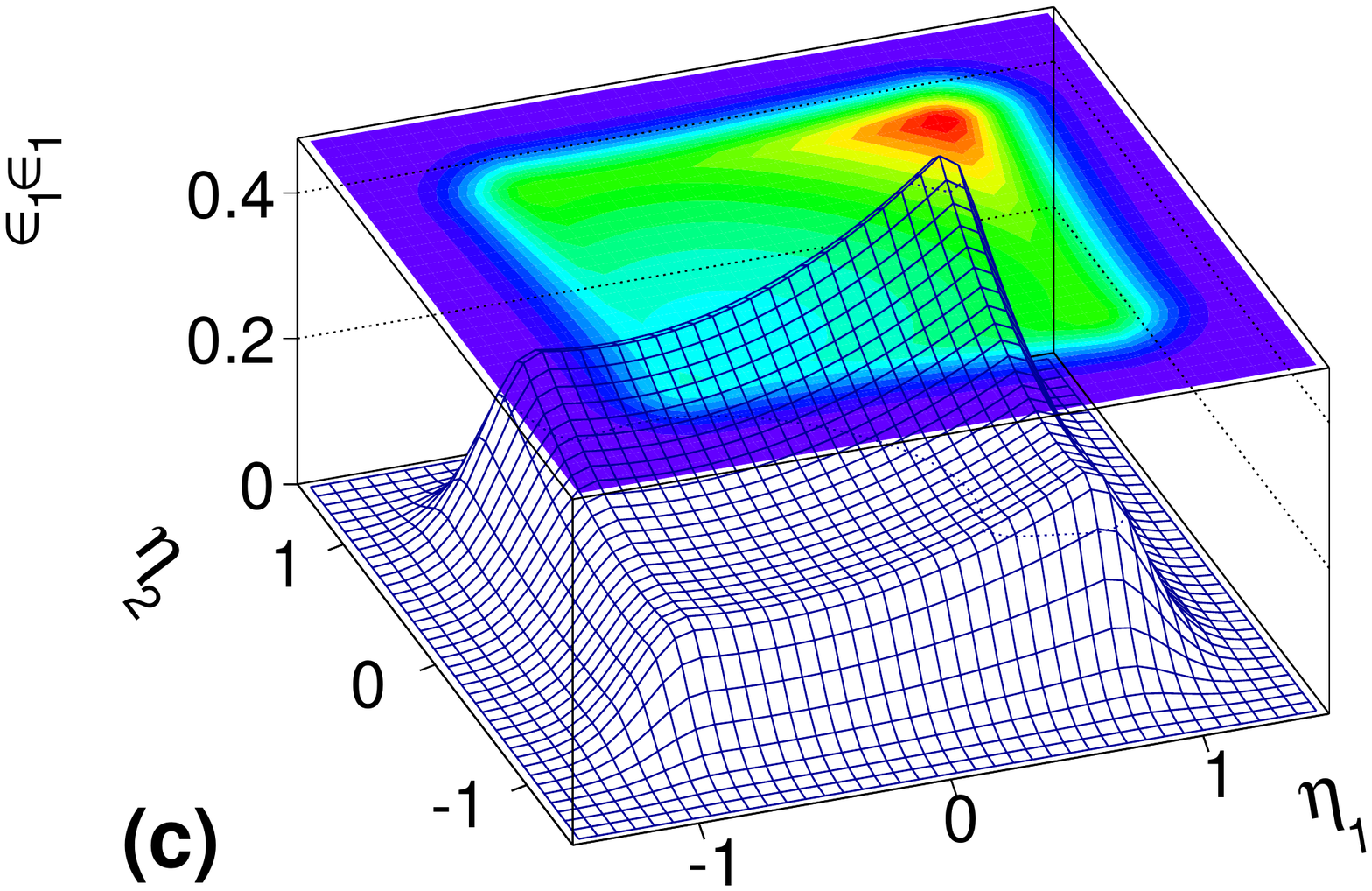}
\caption{(Color Online) Pair detection efficiency dependence on $\eta_1$, $\eta_2$ for (a) perfect efficiency, 
(b) flat response with smooth edges, and non-linear response with edge effects shown for $\epsilon_o=0.7, \alpha=0.4$, $\beta=0.4$. }
\label{fig:BasicPairEfficiencyVsEta}
\end{figure}

\begin{figure}[t!]
\centering
\includegraphics[width=0.32\textwidth]{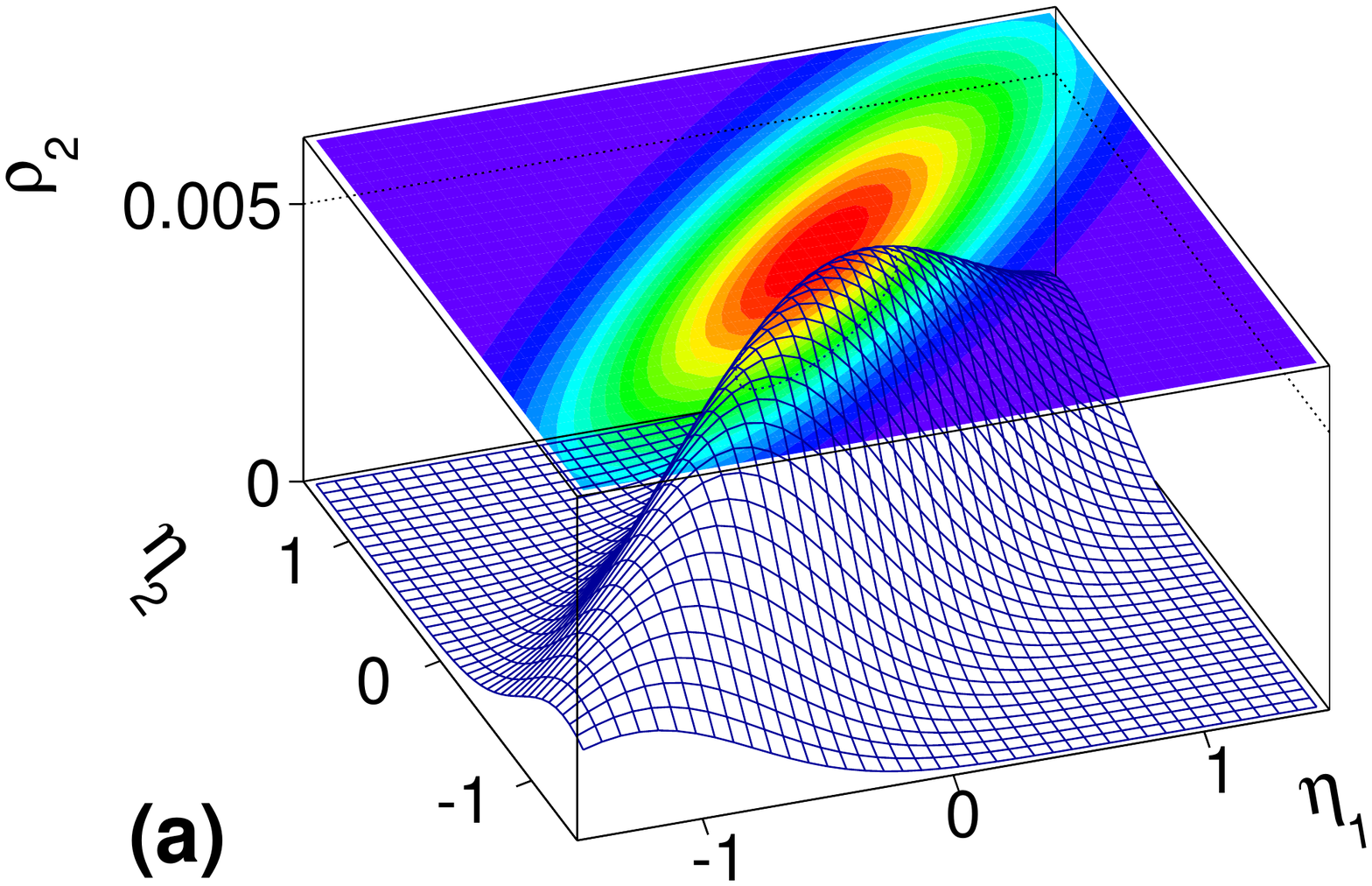}
\includegraphics[width=0.32\textwidth]{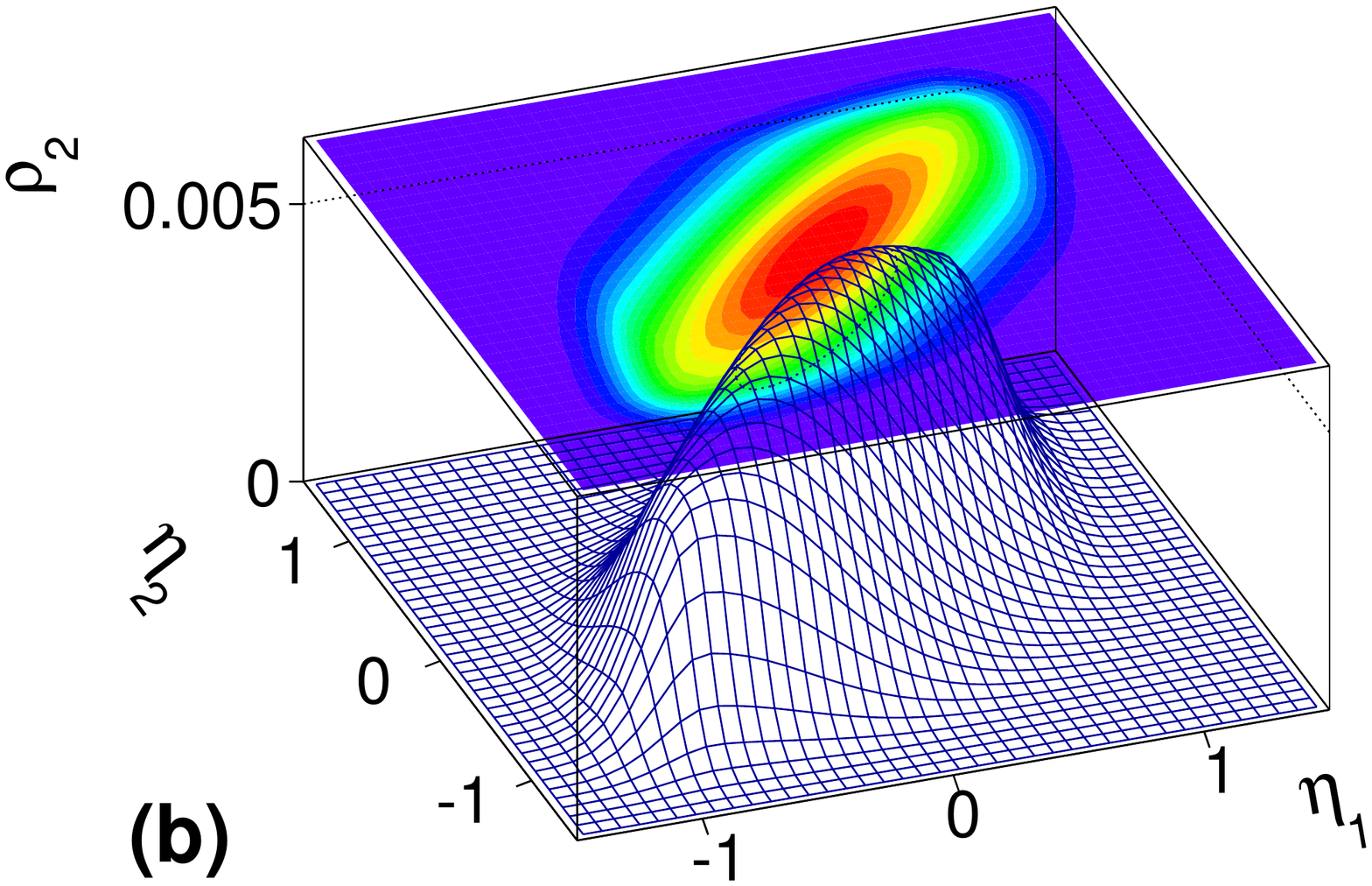}
\includegraphics[width=0.32\textwidth]{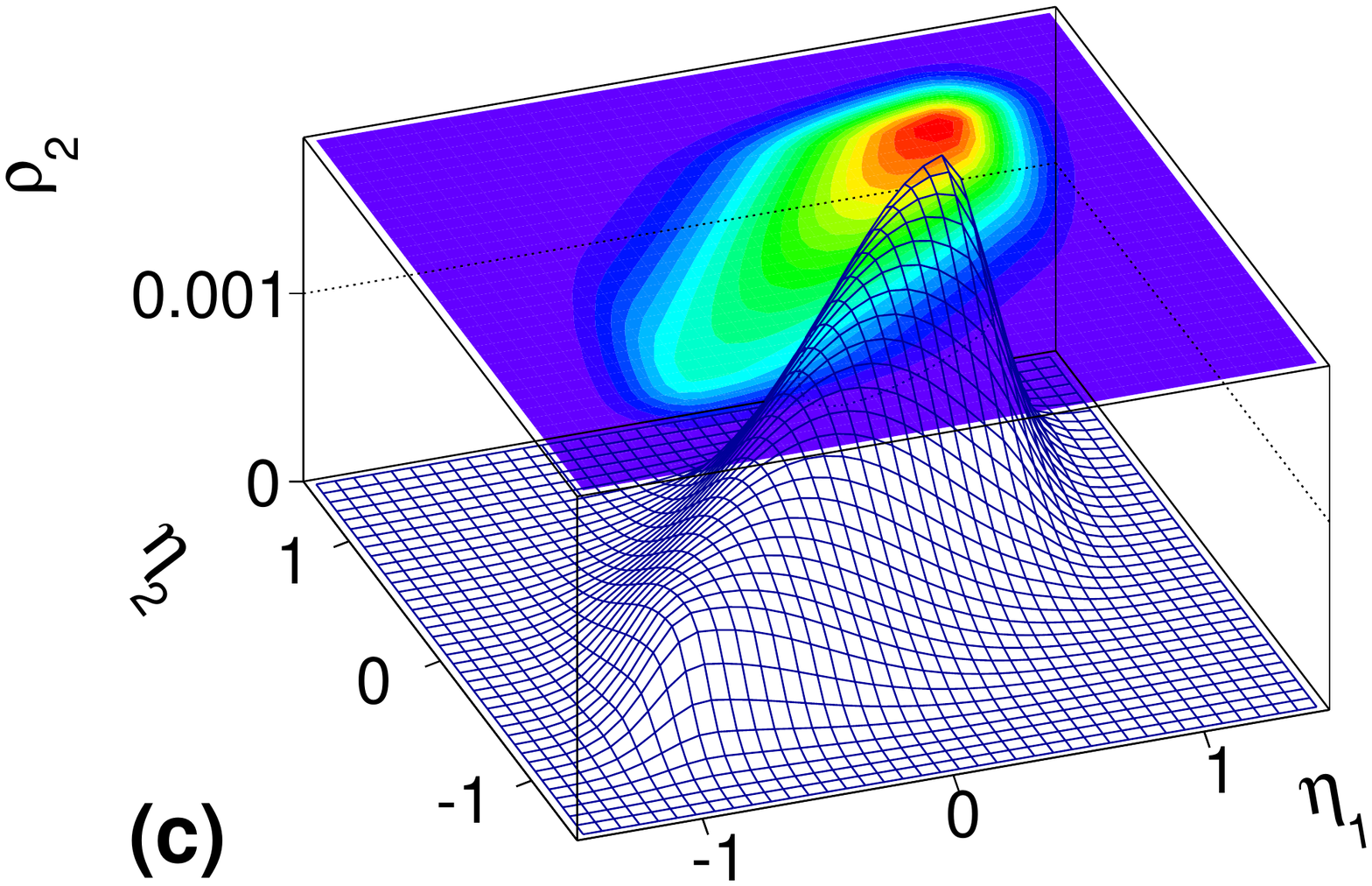}
\caption{(Color Online) Pair yield dependence on $\eta_1$, $\eta_2$ for (a) perfect efficiency, 
(b) flat response with smooth edges, and non-linear response with edge effects shown for $\epsilon_o=0.7, \alpha=0.4$, $\beta=0.4$. }
\label{fig:BasicPairYieldVsEtaVsEta}
\end{figure}
\begin{figure}[t!]
\centering
\includegraphics[width=0.32\textwidth]{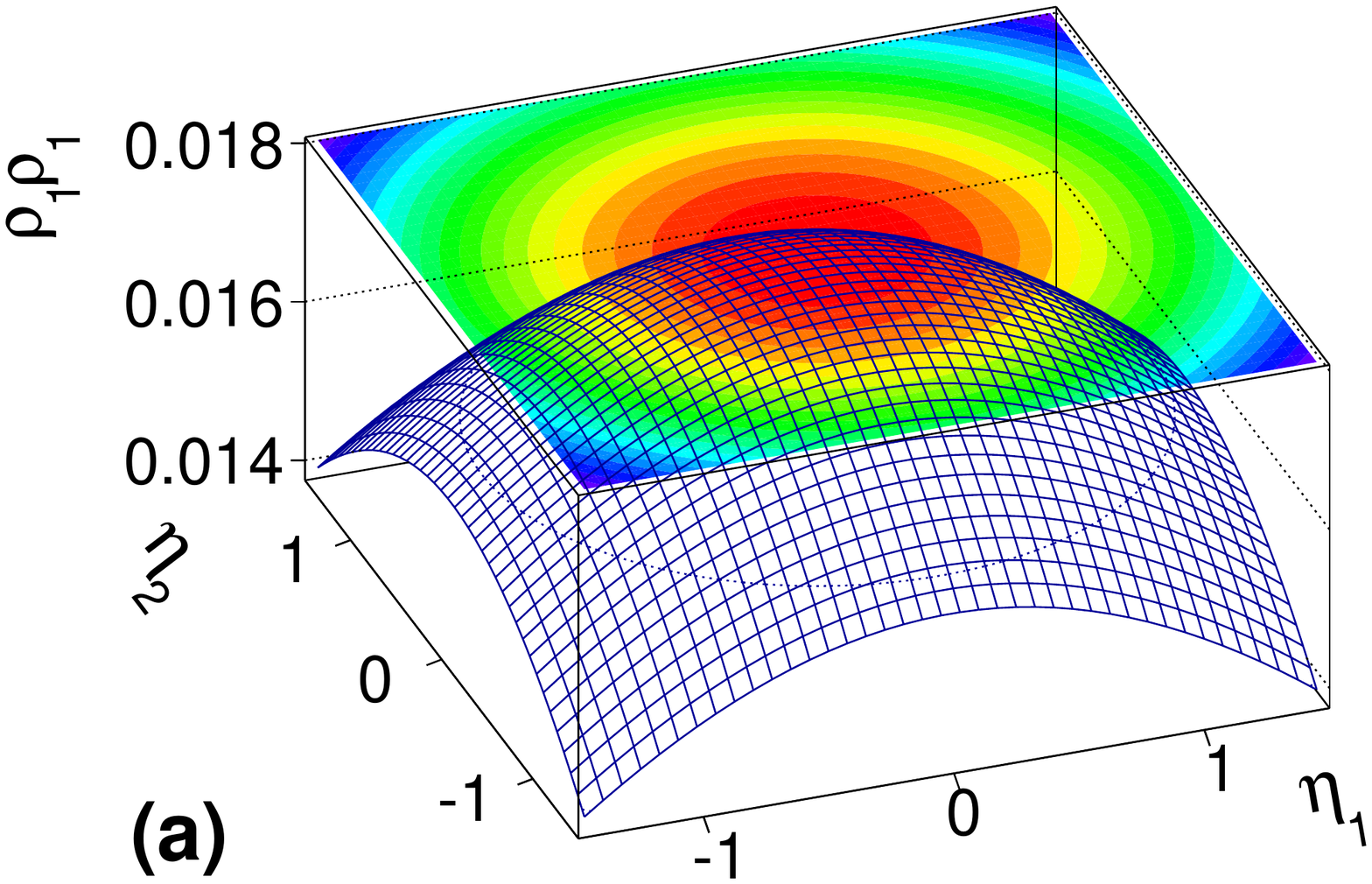}
\includegraphics[width=0.32\textwidth]{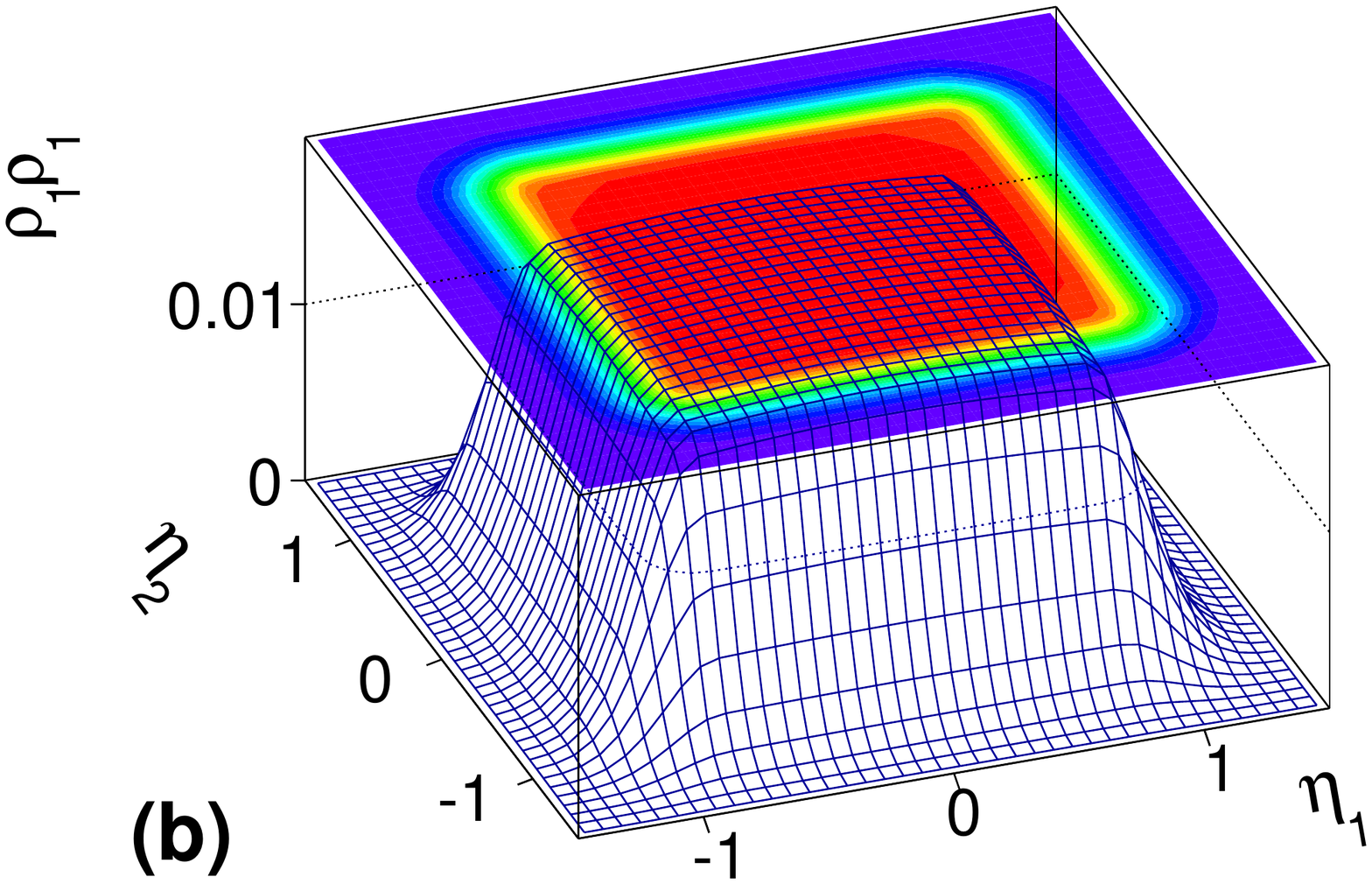}
\includegraphics[width=0.32\textwidth]{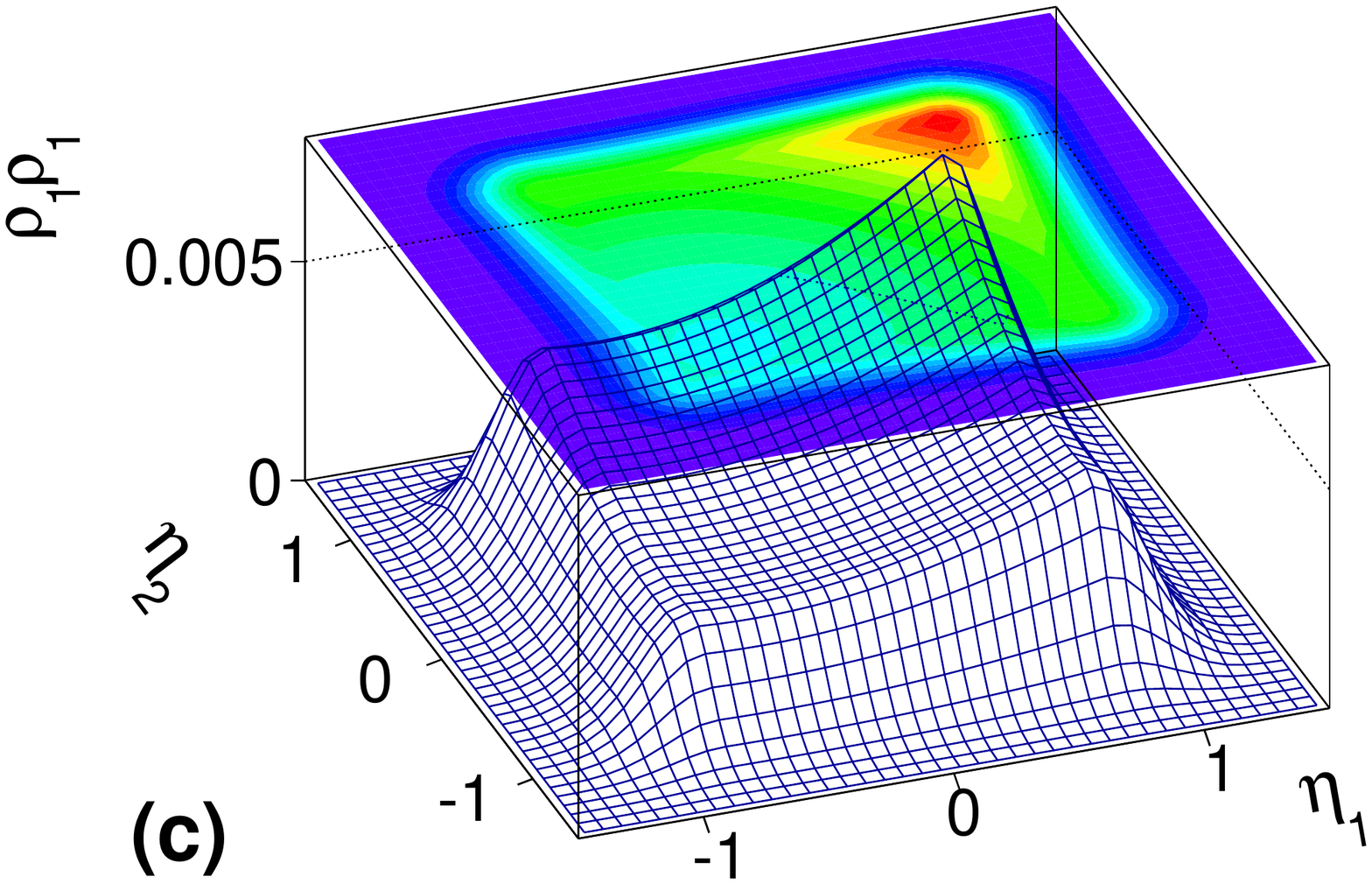}
\caption{(Color Online) Product of singles $\rho_1\rho_1$ vs. $\eta_1$, $\eta_2$ for (a) perfect efficiency, 
(b) flat response with smooth edges, and non-linear response with edge effects shown for $\epsilon_o=0.7, \alpha=0.4$, $\beta=0.4$. }
\label{fig:BasicSinglesProductVsEtaVsEta}
\end{figure}
\begin{figure}[t!]
\centering
\includegraphics[width=0.32\textwidth]{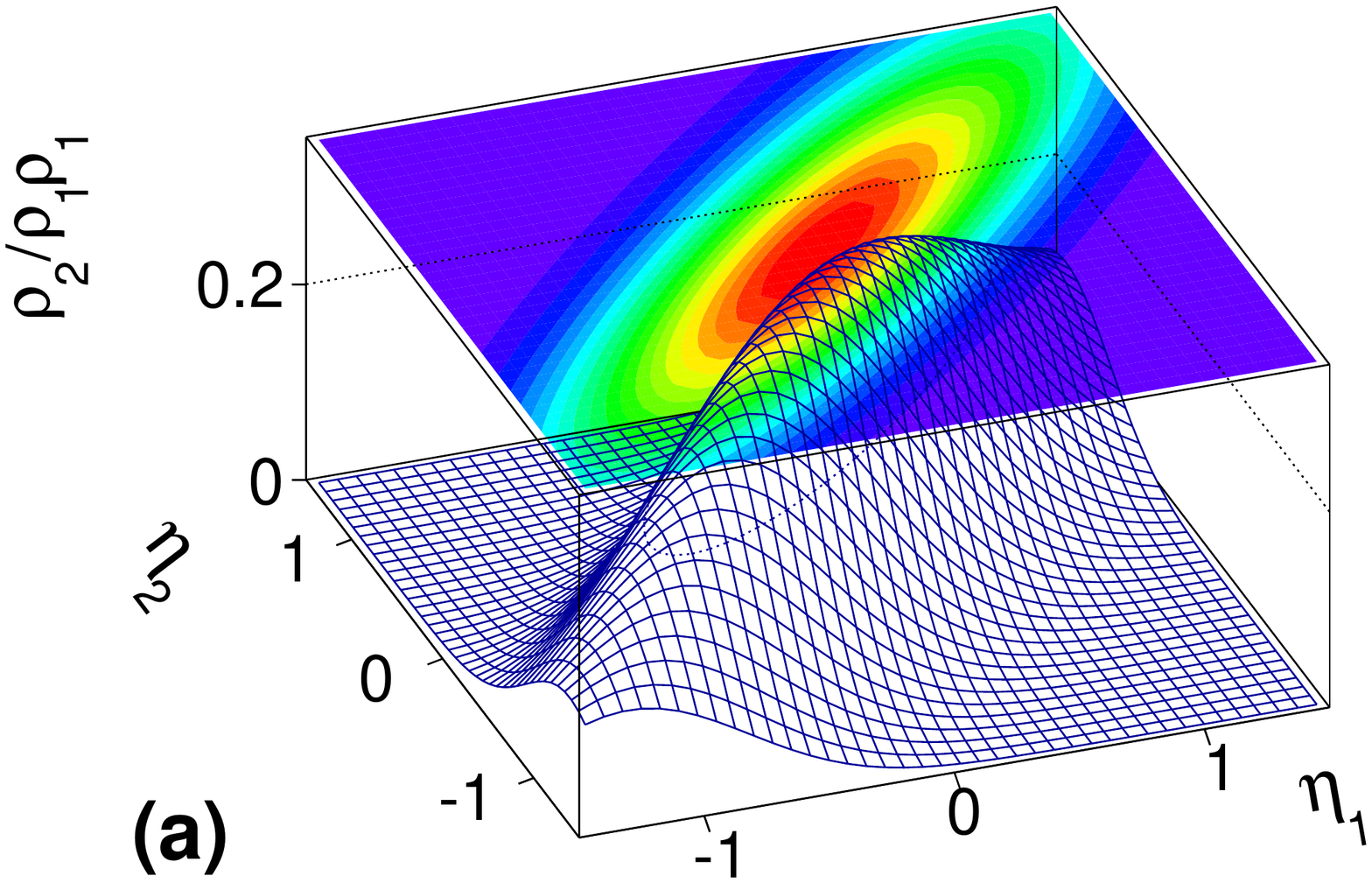}
\includegraphics[width=0.32\textwidth]{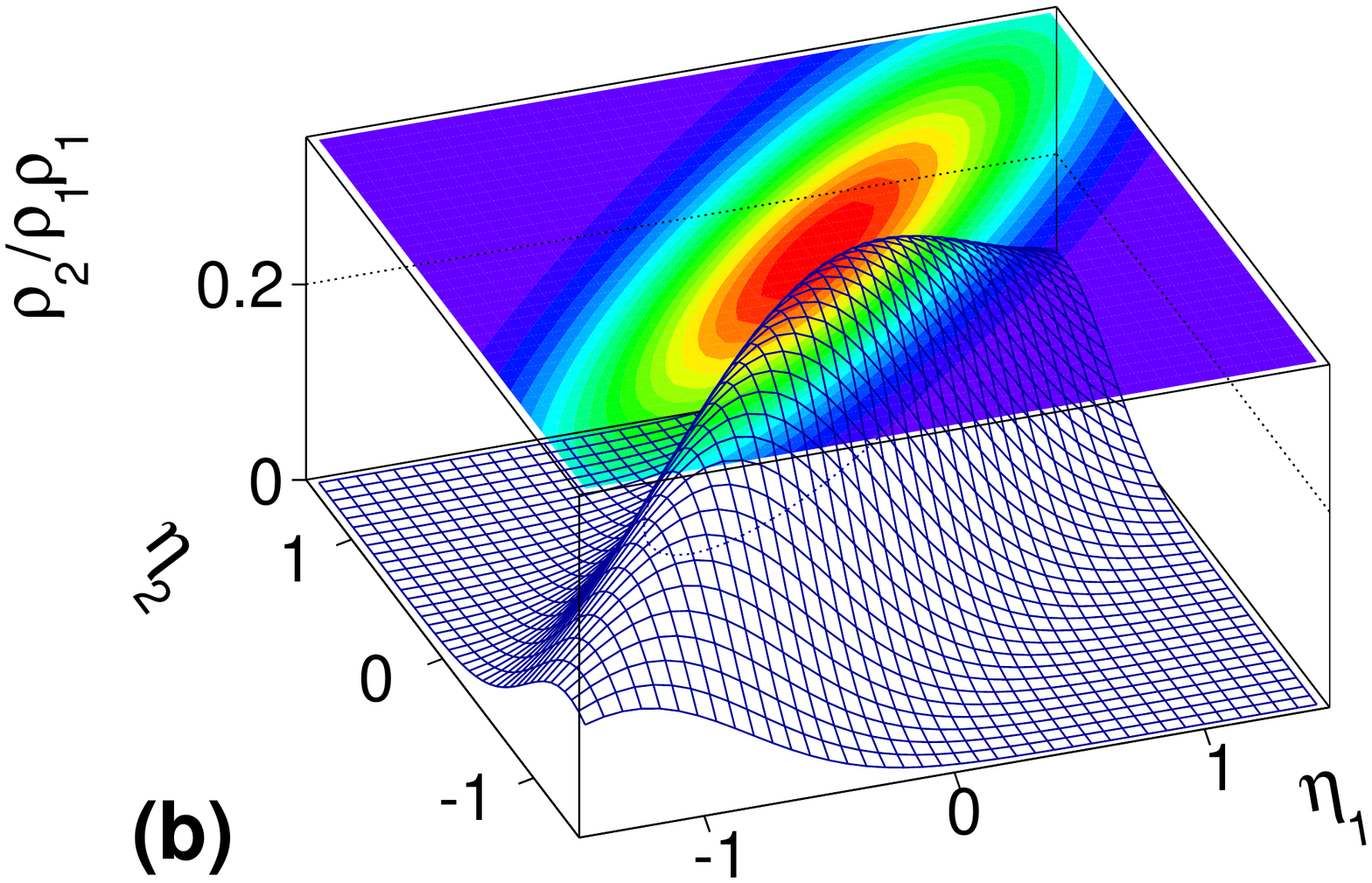}
\includegraphics[width=0.32\textwidth]{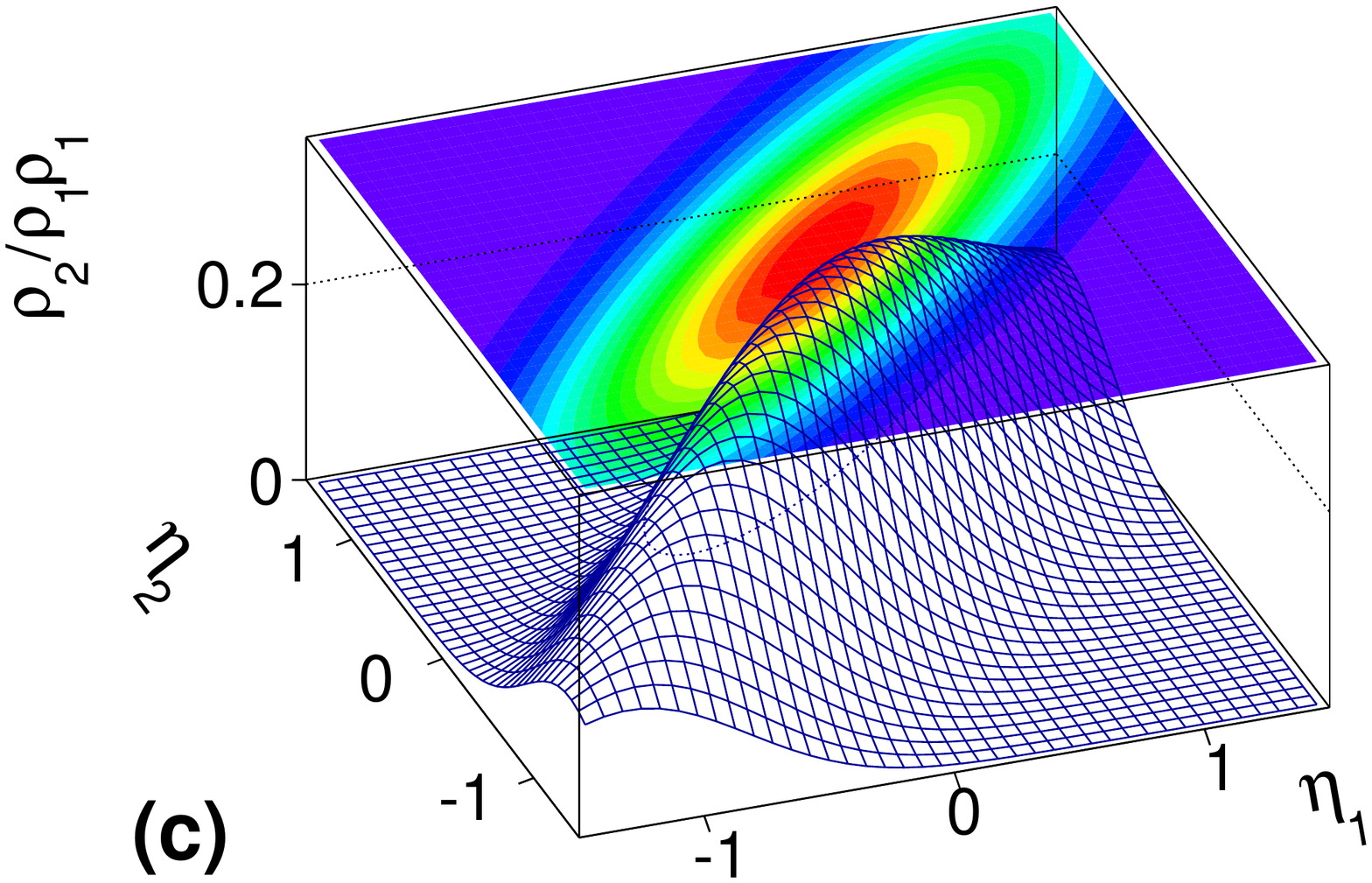}
\caption{(Color Online) Ratio of pair yield to product of single particle yields vs. $\eta_1$, $\eta_2$ for (a) perfect efficiency, 
(b) flat response with smooth edges, and non-linear response with edge effects shown for $\epsilon_o=0.7, \alpha=0.4$, $\beta=0.4$. }
\label{fig:R2M2VsEtaVsEta}
\end{figure}

\begin{figure}[h]
\includegraphics[width=0.48\textwidth]{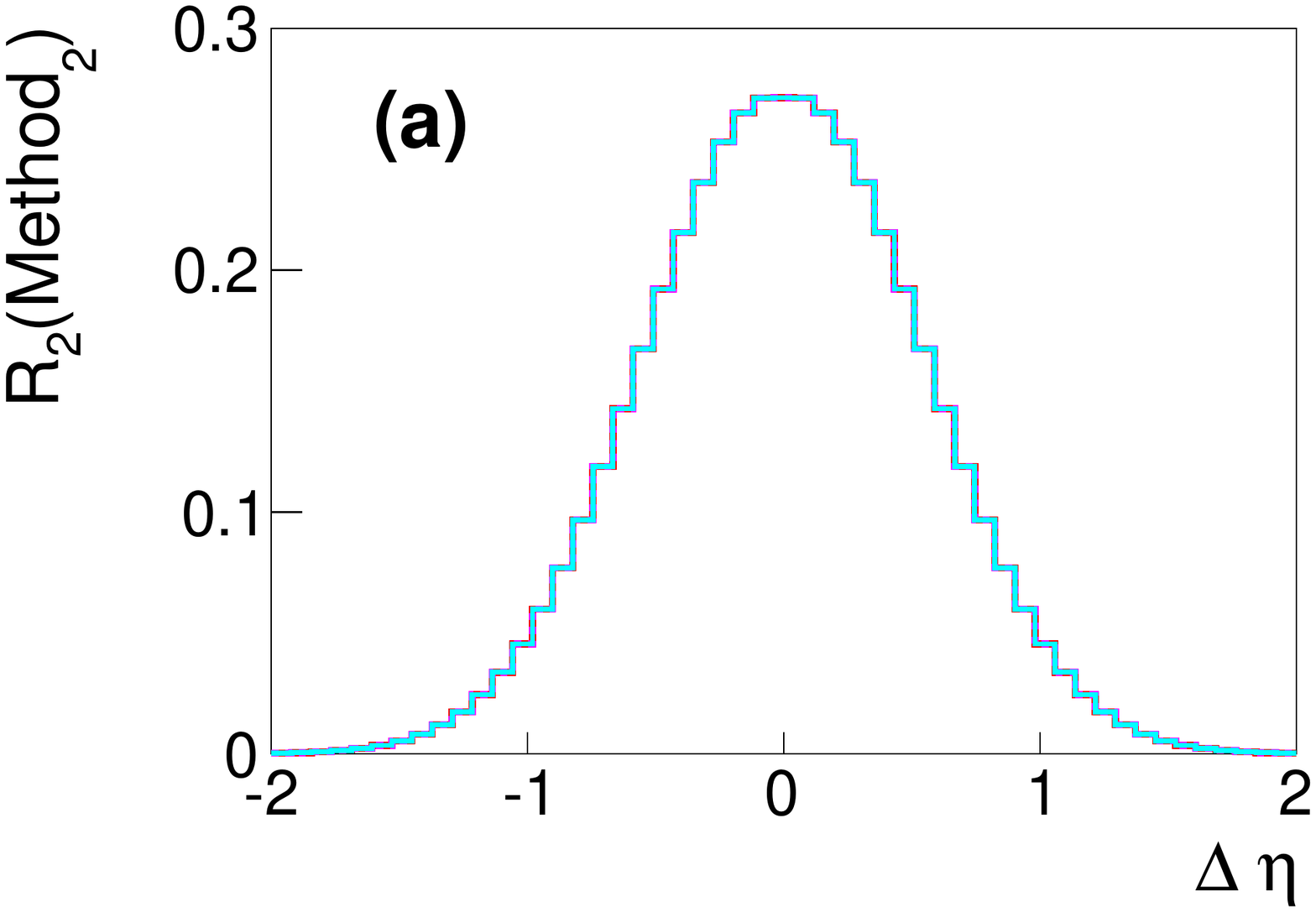}
\includegraphics[width=0.48\textwidth]{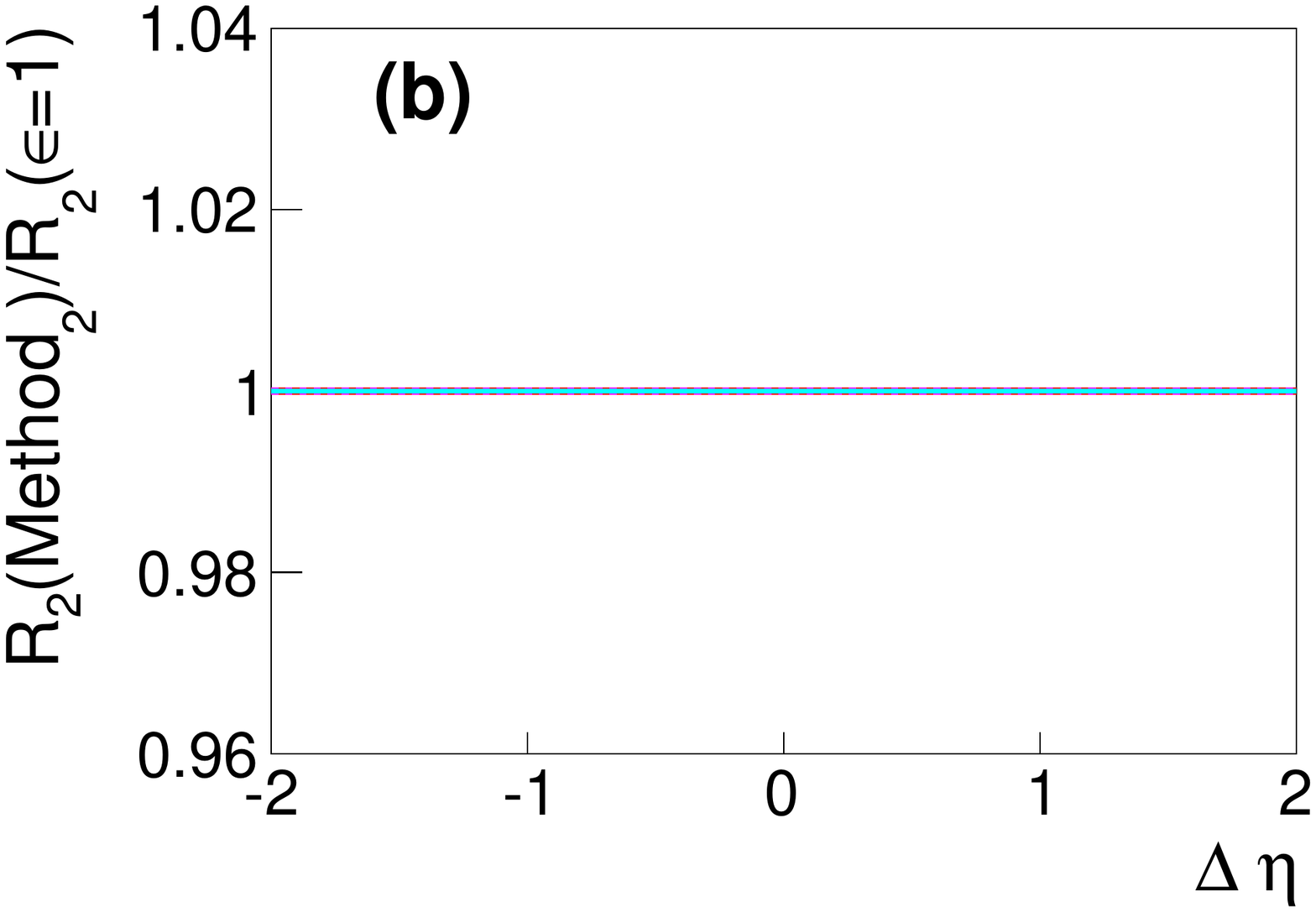}
\caption{(Color Online) (left) Function $R_2(\Delta\eta)$ obtained by $\overline{\eta}$ average of $\rho_2/\rho_1\rho_1$ distributions
obtained with (a) perfect efficiency, b) flat response with smooth edges, and non-linear response with edge effects shown for $\epsilon_o=0.48$, $\alpha=0.4$, $\beta=0.4$.(right) Ratios of the distributions $R_2(\Delta\eta)$ obtained by Method 2 for imperfect efficiency to that obtained with
perfect efficiency.}
\label{fig:BasicR2M2VsDeta}
\end{figure}
\begin{figure}[h]
\includegraphics[width=0.48\textwidth]{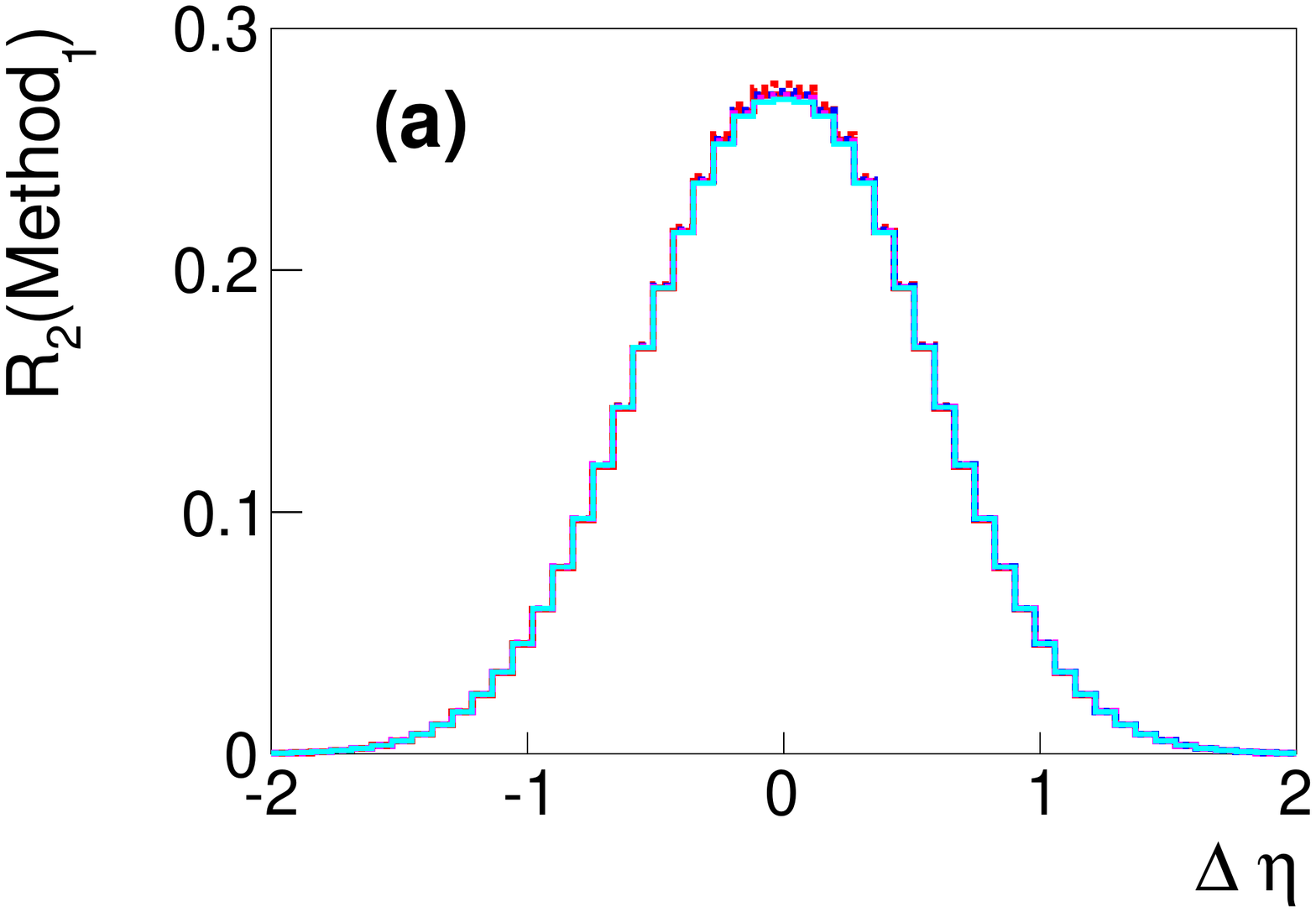}
\includegraphics[width=0.48\textwidth]{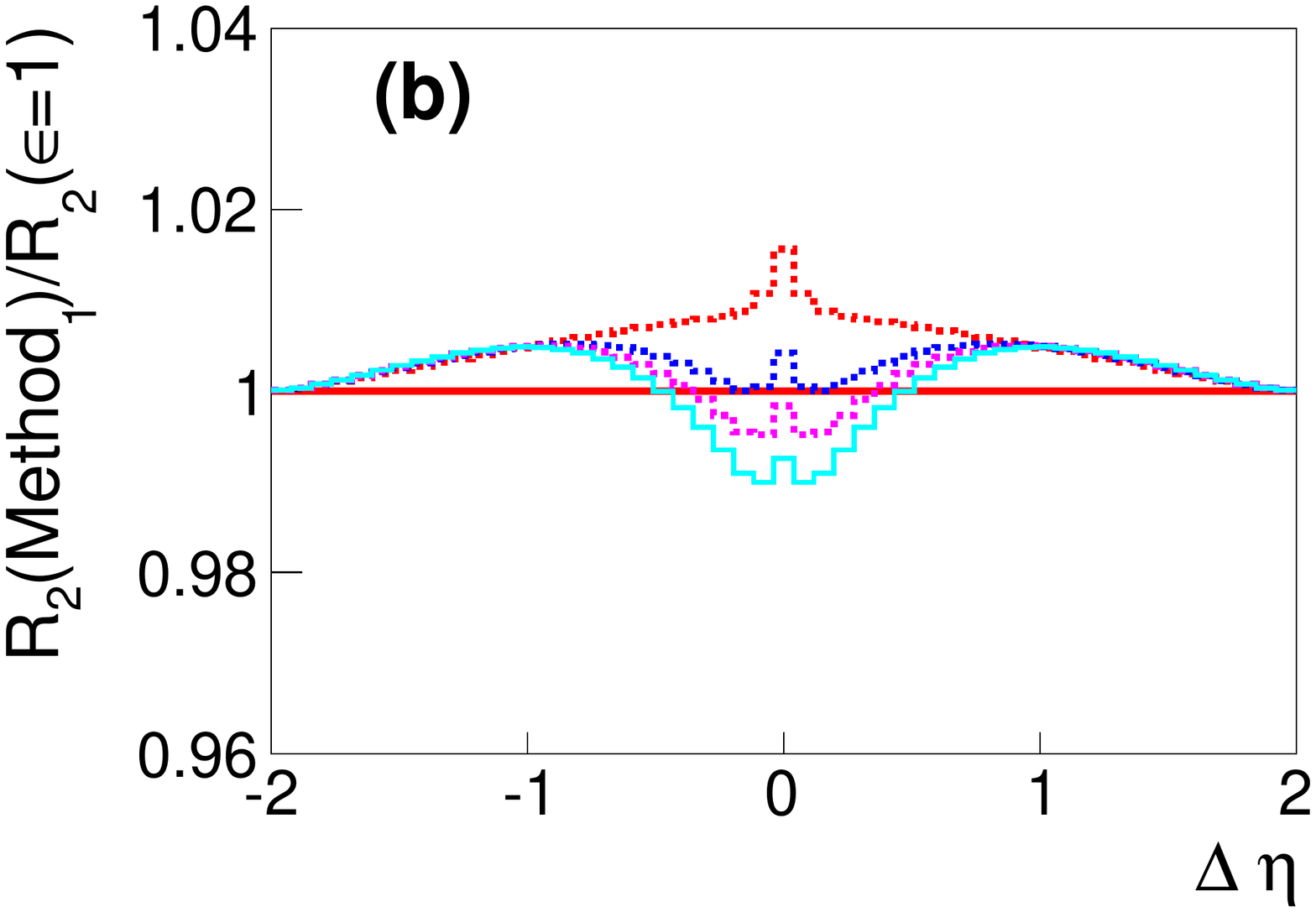}
\caption{(Color Online) (left) Function $R_2(\Delta\eta)$ obtained with Method 1 for
cases of  perfect efficiency,  flat response with smooth edges, and non-linear response with edge effects shown for $\epsilon_o=0.7$, $\alpha=0.2$, $\beta=0.2$ (dash blue),  
$\epsilon_o=0.56$, $\alpha=0.3$, $\beta=0.3$ (dash purple), 
$\epsilon_o=0.48$, $\alpha=0.4$, $\beta=0.4$ (solid light blue).
(right) Ratios of these distributions to that obtained with perfect efficiency.}
\label{fig:BasicR2M1VsDeta}
\end{figure}
\end{widetext}

For illustrative purposes and discussions of efficiency corrections, we introduce a simple correlated particle production model defined as follows. Particle emission is assumed to have a Gaussian dependence on pseudo rapidity.
\begin{eqnarray} 
\rho_1(\eta) \propto  \exp(-\frac{\eta^2}{2\sigma_{\eta}^2}) 
\end{eqnarray} 
where $\sigma_{\eta}$ is the (single particle) width of the pseudo rapidity distribution.
We  additionally model two particle correlations with an ad hoc Gaussian 
dependence.
\begin{eqnarray} 
\rho_2(\eta_1,\eta_2) \propto  1 + C \exp(-\frac{\Delta \eta^2}{2 \sigma_{\Delta\eta}^2}) \exp(-\frac{\overline{\eta}^2}{2\sigma_{\overline{\eta}}^2})
\end{eqnarray} 
with
\begin{eqnarray} 
\Delta \eta &=& \eta_1 - \eta_2 \\ \nonumber
\overline{\eta} &=& (\eta_1 + \eta_2)/2
\end{eqnarray} 
The coefficient $C$ expresses the maximum strength of the correlation while $\sigma_{\Delta\eta}$ and $\sigma_{\overline{\eta}}$ determine the width of the correlation function relative to the pseudo rapidity difference, $\Delta\eta$, and average pseudo rapidity of the pair, $\overline{\eta}$.

We use the model  with $\sigma_{\eta}= 3$, $C=1$, $\sigma_{\Delta\eta}=0.5$, and $\sigma_{\overline{\eta}}=2$, to illustrate the impact of the detection efficiency on the measured particle pair yield in Fig. \ref{fig:BasicPairYieldVsEtaVsEta}.  Clearly, the non-uniform efficiency response of the
detection system may considerably modify the strength of the measured two-particle yield. Method 2 however corrects for detection effects by dividing 
the pair yield by the product of singles $\rho_1 \times \rho_1$ obtained either by a mixed event technique or by multiplying the singles spectra onto itself. 
The product of singles $\rho_1 \times \rho_1$ corresponding to the same three cases are shown in Fig. \ref{fig:BasicSinglesProductVsEtaVsEta}.  
They are used to divide out the efficiencies and obtain the pair to single yield ratio $\rho_2/\rho_1\rho_1$ displayed in Fig. \ref{fig:R2M2VsEtaVsEta}. One verifies
by direction inspection that the three distributions are identical as expected from the definition of Method 2 and the assumed factorization of efficiency 
used in generating the plots.  Fig. \ref{fig:BasicR2M2VsDeta} (a) presents the $\overline{\eta}$ averaged distributions, $R_2(\Delta\eta)$, obtained by 
calculations with perfect efficiency, flat response with smooth edges, and non-linear response with edge effects shown for $\epsilon_o=0.7, \alpha=0.4$, $\beta=0.4$.
Fig. \ref{fig:BasicR2M2VsDeta} (b) displays the ratio of distributions obtained with imperfect efficiency to that obtained for perfect efficiency and illustrates 
that all distributions are virtually identical and perfectly corrected by Method 2 as expected.

Correlation analyses are however routinely carried out with Method 1 rather than Method 2. Given that Method 1 uses a ratio of pair yield and product of singles, one might expect it to produce robust correlation functions. We proceed 
to show that while Method 1 yields results that are approximately robust, it can in fact produce correlation function that arbitrarily deviate from the actual distribution. Rather 
than calculate the ratio of pair yields and product of single particle yields as a function $\eta_1$ and $\eta_2$, Method 1 uses yields calculated explicitly  as a function of 
$\Delta\eta$. The $\overline{\eta}$ averaging is carried out independently for pairs and product of singles (or mixed events).  The 
ratio $R_2(\Delta\eta)$ measured with method 1 can be formally written as follows. 
\begin{eqnarray} 
R_2(\Delta\eta)^{Method 1} = \frac{\int g(\Delta\eta,\overline{\eta})R_2^{true}(\Delta\eta,\overline{\eta})d\overline{\eta}}{\int g(\Delta\eta,\overline{\eta})d\overline{\eta}} 
\end{eqnarray} 
where $R_2^{true}$ is the true value of the correlation function, and $g(\Delta\eta,\overline{\eta})= \epsilon_1 \times \epsilon_1 \times \rho_1 \times \rho_1(\Delta\eta,\overline{\eta})$ is, in general, a non trivial function of $ \Delta\eta,\overline{\eta}$. Obviously, the function $g$  cannot be factored out of the integrals. Method 1 is consequently intrinsically non-robust. This unfortunate conclusion is illustrated with
the simple correlation model introduced above.  Figure \ref{fig:BasicR2M1VsDeta} (left)  displays functions $R_2(\Delta\eta)$
obtained with Method 1 for perfect efficiency (solid red), a flat response with smooth edges (dash red), 
and non-linear response with edge effects shown for 
$\epsilon_o=0.7$, $\alpha=0.2$, $\beta=0.2$ (dash blue),  
$\epsilon_o=0.56$, $\alpha=0.3$, $\beta=0.3$ (dash purple), 
$\epsilon_o=0.48$, $\alpha=0.4$, $\beta=0.4$ (solid light blue). One finds that the distributions are remarkably similar in spite of the large differences 
of efficiency used in their calculations. Differences exist  however and 
are easily visualized in Fig. \ref{fig:BasicR2M1VsDeta} (right) from the ratios of distributions to that obtained for perfect efficiency. Differences are maximum 
near $\Delta\eta \sim 0$ and amount to deviations of a few percent only. Note that the magnitude of the deviation strongly depends 
on the $\overline{\eta}$ dependence of the correlation function. If the dependence is weak, or the correlation function essentially constant
within the $\overline{\eta}$ acceptance, than deviations are very small. However, if both the efficiency and the correlation function
exhibit rapid dependence on $\overline{\eta}$, than arbitrarily large deviation can occur between the measured 
and actual correlation function. We thus conclude that Method 1 is non robust for measurements of correlations as a function of $\Delta\eta$. 
It however may remain reliable and sufficient in a wide variety of contexts and analyses, provided of course, as for Method 2, the pair efficiency 
factorizes. Wherever high accuracy is required and strong variations of the efficiency throughout the acceptance are present, Method 2 is however strongly advised.

There are  several instrumentals effects that may break the factorization. We discuss three such effects in  sections \ref{sec:efficiencyVsZ},  \ref{sec:efficiencyVsMult} and \ref{sec:efficiencyVsL}. However, we first discuss the case of correlation functions measured as a function of the
difference between two particles azimuthal angles, $\Delta\phi$, for which periodic boundary conditions lead to considerable simplifications and robustness of 
correlation functions obtained with Method 1 as well as Method 2.

\subsection{Azimuthal Distributions}
\label{sec:efficiencyVsDphi}

Correlation functions measured as a function of the relative azimuthal angle of emission of two particles constitute a special case where both Method 1 and Method 2 are robust in contrast to the longitudinal correlation functions considered in the previous section where only Method 2 is strictly robust. The difference stems from periodic boundary conditions that apply to azimuthal distributions but obviously not to longitudinal correlation functions. To verify this assertion, we describe the detection efficiency in terms of a generic Fourier decomposition, $\epsilon(\phi)$, as follows.
\begin{eqnarray} 
\epsilon(\phi_i) = \epsilon_o [1+\sum_m a_m \cos(m(\phi_i-\alpha_m))]
\end{eqnarray} 
where $\phi_i$, with $i=1, 2$ is the azimuthal angle of particles, $\epsilon_o$ is the average efficiency, $a_m$  are Fourier coefficients,  and  $\alpha_m$ are phases.
Likewise, the pair yield $\rho_2(\Delta\phi)$ can also be expressed in terms of a Fourier decomposition.
\begin{eqnarray} 
\rho_2(\Delta\phi) = \rho_o^2 [1+\sum_n b_n \cos(n(\phi_1-\phi_2))]
\end{eqnarray} 
where $\rho_o$ represents the single particle yield, and $b_n$ are Fourier coefficients chosen to properly model the correlation function. The measured
number of pairs is thus
\begin{widetext}
\begin{eqnarray} 
n_2(\Delta\phi) &= &\epsilon_o^2 \rho_o^2 \int_o^{2\pi} \int_o^{2\pi} [1+\sum_m a_m \cos(m(\phi_1-\alpha_m))][1+\sum_p a_p \cos(p(\phi_2-\alpha_p))]
\\ \nonumber
 & & \times [1+\sum_n b_n \cos(n(\phi_1-\phi_2))] \delta(\Delta\phi - \phi_1 +\phi_2) d\phi_1 d\phi_2
\end{eqnarray} 
\end{widetext}
The product of singles $n_1\times n_1(\Delta\phi)$ is similarly given by 
\begin{widetext}
\begin{eqnarray} 
n_1\times n_1(\Delta\phi) &= &\epsilon_o^2 \rho_o^2 \int_o^{2\pi} \int_o^{2\pi} [1+\sum_m a_m \cos(m(\phi_1-\alpha_m))][1+\sum_p a_p \cos(p(\phi_2-\alpha_p))]
\\ \nonumber
 & & \times \delta(\Delta\phi - \phi_1 +\phi_2) d\phi_1 d\phi_2
\end{eqnarray} 
\end{widetext}
It is straightforward to show that the ratio $n_2/n_1n_1$ simplifies and is independent of efficiencies owing to the periodicity of the efficiency function, $\epsilon(\phi)$.
\begin{eqnarray} 
\frac{n_2}{n_1n_1}(\Delta\phi) &=& 1+\sum_n b_n \cos(n(\Delta\phi))\\ \nonumber
  & = & \frac{\rho_2}{\rho_1 \rho_1}(\Delta\phi)
\end{eqnarray} 
Method 1, as well as Method 2,  thus  yields a robust measurement of azimuthal correlation functions. The obvious lack of periodicity of $\epsilon(\eta)$ however forbids this convenient simplification for measurement
of longitudinal correlation functions and Method 1 is not strictly robust in that case.

\section{Efficiency dependence on the collision position}
\label{sec:efficiencyVsZ}

\begin{figure}[h]
\includegraphics[width=0.45\textwidth]{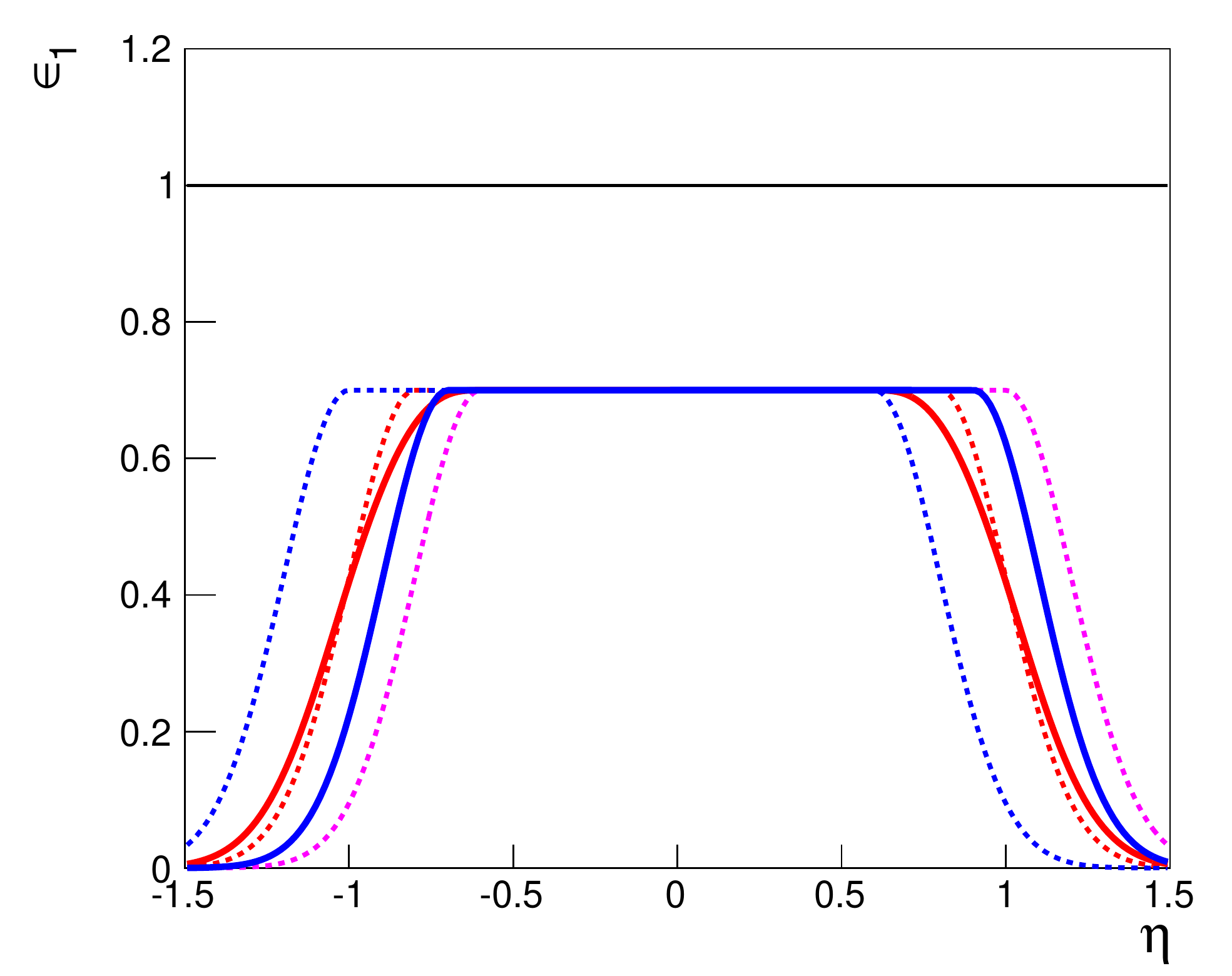}
\caption{(Color Online) Model of the  vertex position, $z$ dependence of the detection efficiency used in sec.\ref{sec:efficiencyVsZ}. See text for description of the model.}
\label{fig:Epsilon1VsEtaVsZ}
\end{figure}

At colliders, the longitudinal extent of the beam-beam interaction  region has a finite size determined by the optics of the beams. Given the geometry of the detectors, this implies that the effective boundaries beyond which the efficiency  rapidly drops to zero actually depend on the position of the interaction vertex as illustrated in Fig. \ref{fig:Epsilon1VsEtaVsZ}. Effectively, the efficiency becomes a function of both $\eta$ and the position, $z$,  of the interaction vertex, which we note $\epsilon(\eta|z)$. Given the position of the interaction vertex is a random variable, the detection efficiency also becomes a random variable. The "well behaved detector" assumption posited in the previous sections is thus implicitly violated in practice. Though this may have a rather limited impact on measurements of single particle spectra, these "fluctuations" of the efficiency may have devastating effects on correlation function measurements, particularly those carried out as a function of the difference $\Delta\eta$. 
To illustrate this point, we consider the number of particles detected near the edges of the acceptance ($ \eta \sim \pm\eta_1$) in any given event. If the event is produced near $z=0$, then one expects the efficiency at $\eta= \pm \eta_o$ to have some nominal value, of order $\epsilon_o/2$. But if the event is produced at large negative $z$, the acceptance at positive $\eta$ is somewhat larger, and the efficiency is consequently larger than the nominal value. On the other hand, if the event is produced at large positive $z$, then the efficiency at $\eta= \eta_o$ is likely far reduced and may even vanish. The efficiency of detection, particularly at the detector's edges, is consequently correlated with the vertex position. The number of pairs detected at pseudo rapidity difference of the order of $2\eta_o$ is therefore likely to fluctuate dramatically event to event depending on the longitudinal position, $z$, of the event primary vertex. Additionally, since pairs of particles with $\Delta \eta \sim 0$ may be produced near the edge of the detector's acceptance as well as in its center, the number of pairs with $\Delta \eta \sim 0$ is consequently also influenced by fluctuations of the efficiency and will exhibit an explicit dependence on the vertex position. One can describe these "fluctuations" formally if one assumes that for each value of $z$, the two-particle (joint) detection efficiencies can be factorized.
\begin{eqnarray} 
\epsilon_{pair}(\eta_1, \eta_2| z) = \epsilon_1( \eta_1| z) \times \epsilon_1(\eta_2| z)
\end{eqnarray} 
We will further assume that the probability of observing a collision  at a given position $z$ can be described with some probability density, $P_c(z)$. The average number of singles and pairs measured at given values of $\eta_1$ and $\eta_2$ are thus also functions of $z$.
\begin{widetext}
\begin{eqnarray} 
\la n_1(\eta_1) \ra            &=& K \int_{z_{min}}^{z_{max}} P_c(z) \epsilon(\eta_1|z) \la N_1(\eta_1) \ra dz = \la N_1(\eta_1)\ra  f_1(\eta_1) \\  \nonumber
\la n_2(\eta_1,\eta_2) \ra &=& K \int_{z_{min}}^{z_{max}} P_c(z) \epsilon(\eta_1|z) \times \epsilon(\eta_2|z) \la N_2(\eta_1,\eta_2) \ra dz =  \la N_2(\eta_1,\eta_2) \ra f_2(\eta_1,\eta_2)
\end{eqnarray} 
\end{widetext}
with 
\begin{eqnarray} 
  f_1(\eta_1) &=& K\int_{z_{min}}^{z_{max}} P_c(z) \epsilon(\eta_1|z)dz \\ \nonumber
  f_2(\eta_1,\eta_2) &=&K\int_{z_{min}}^{z_{max}} P_c(z) \epsilon(\eta_1|z) \times \epsilon(\eta_2|z)dz
\end{eqnarray} 
where $K$ is a normalization constant determined by the lower, $z_{min}$, and upper, $z_{max}$,  cuts on the interaction vertex, $z$.
\begin{eqnarray} 
K^{-1} = \int_{z_{min}}^{z_{max}} P_c(z) dz
\end{eqnarray} 
The fact that $f_2(\eta_1,\eta_2)$ cannot be factorized as a product $f_1(\eta_1)\times f_1(\eta_2)$
implies the correlation function $R_2(\eta_1,\eta_2)$ is  no longer robust since the efficiencies are intrinsic functions of the interaction position and therefore do not cancel out in the ratio. 
\begin{widetext}
\begin{eqnarray} 
R_2(\eta_1,\eta_2) =\frac{f_2(\eta_1,\eta_2)}{f_1(\eta_1)f_1(\eta_2)} \frac{ \la N_2(\eta_1,\eta_2) \ra }{ \la N_1(\eta_1) \ra \la N_1(\eta_2) \ra }
\end{eqnarray} 

\begin{figure}[tb!]
\centering
\includegraphics[width=0.32\textwidth]{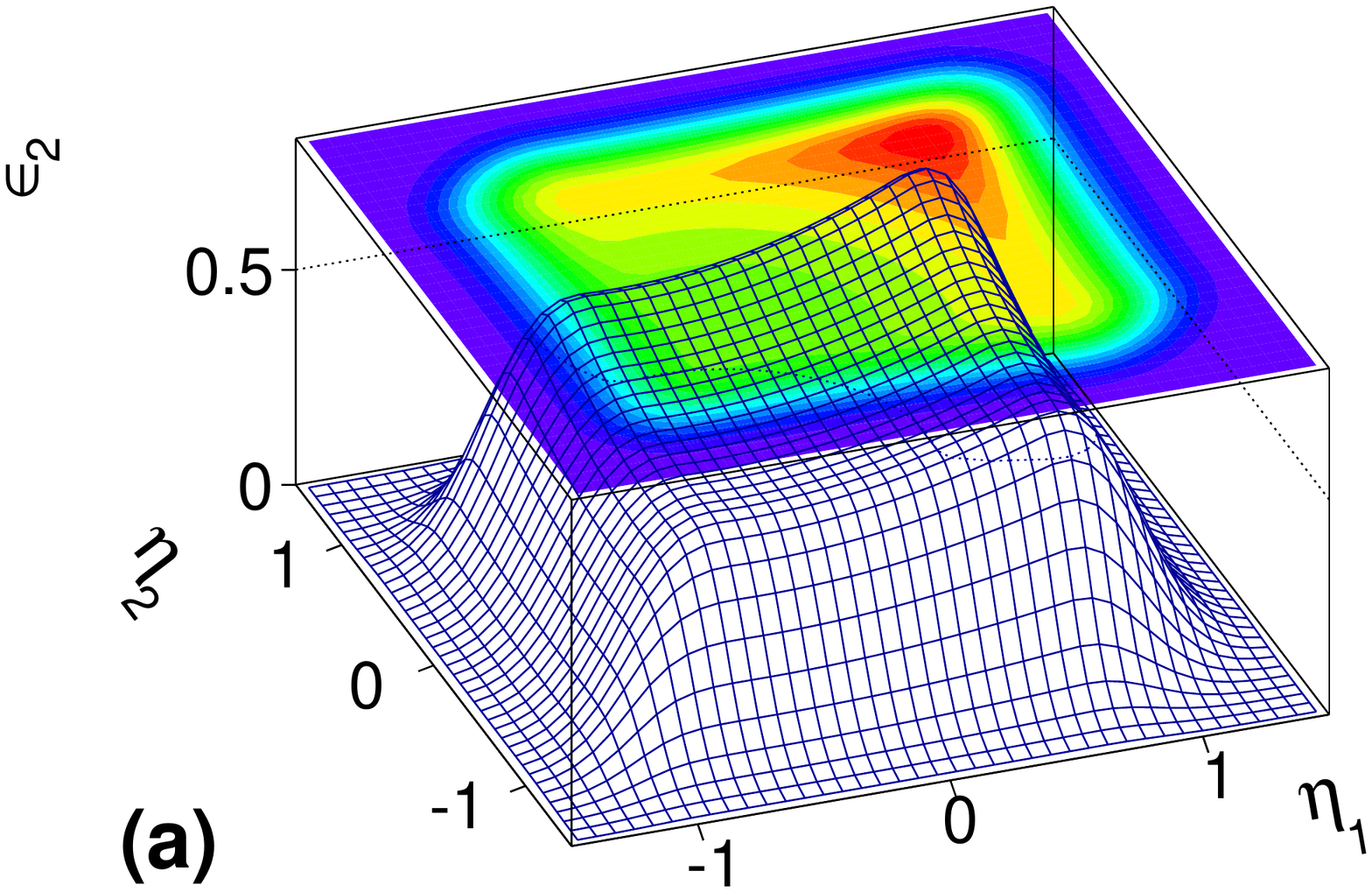}
\includegraphics[width=0.32\textwidth]{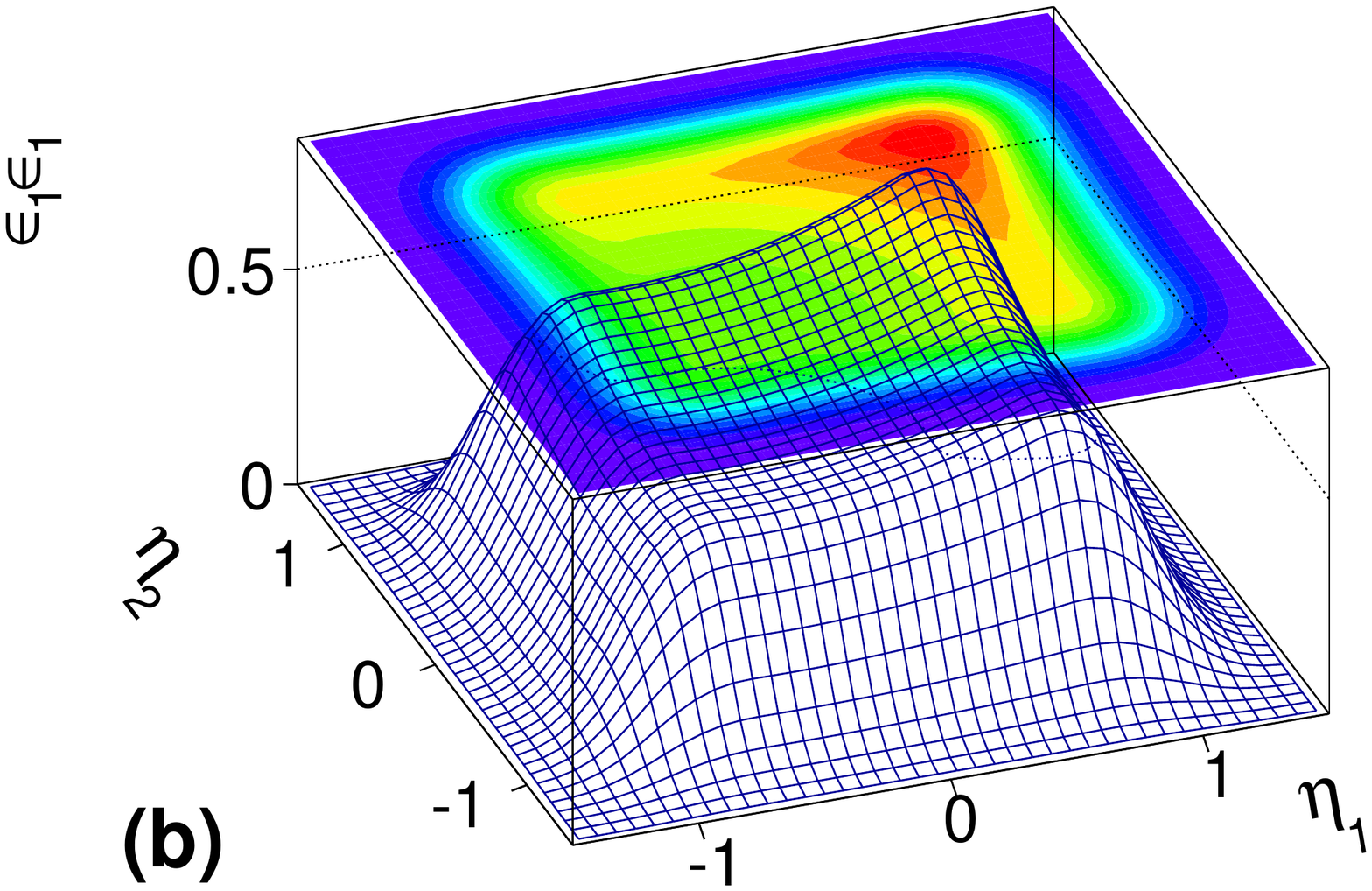}
\includegraphics[width=0.32\textwidth]{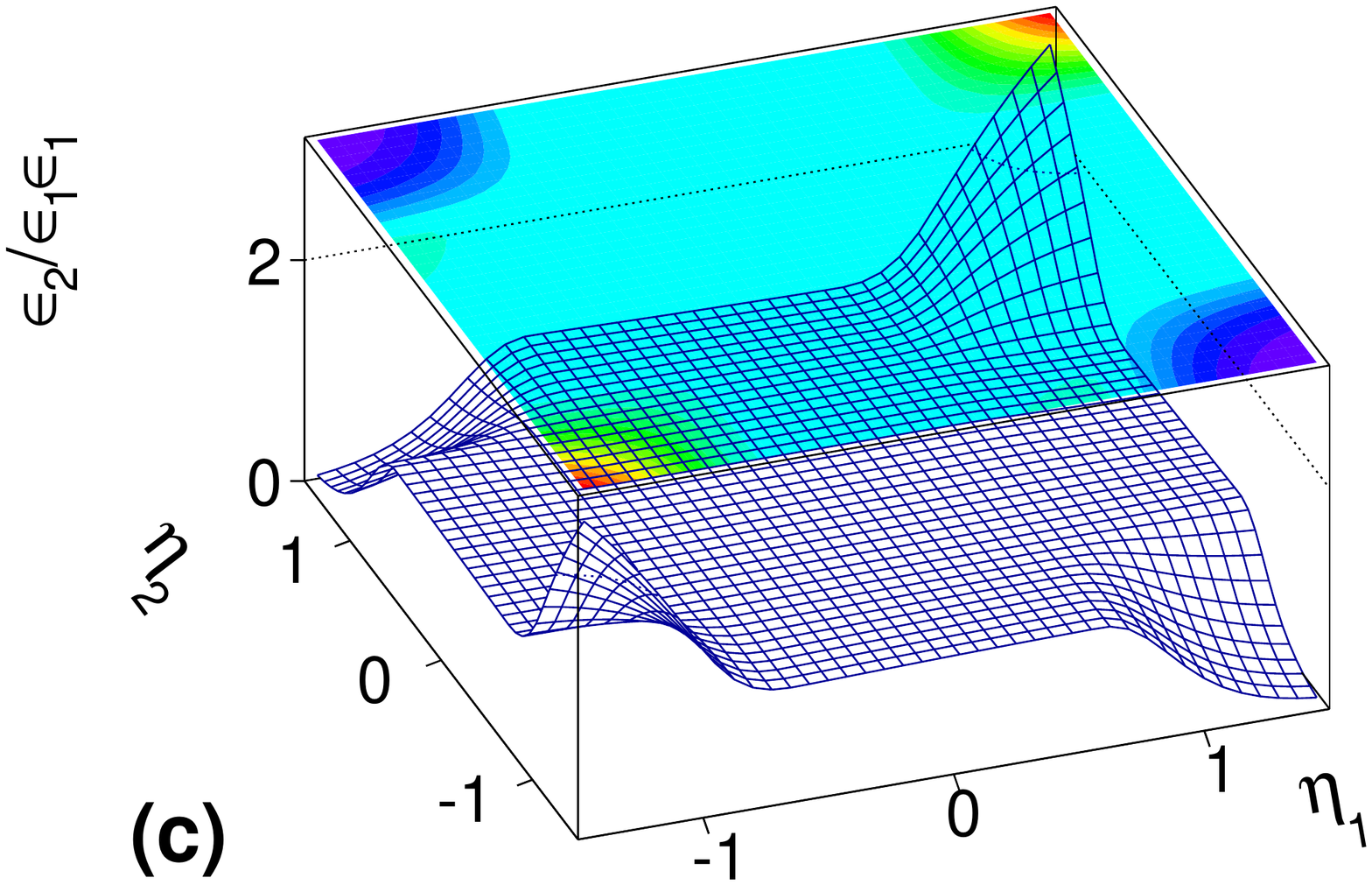}
\caption{(Color Online) (a) Average pair efficiency $\epsilon_2 = f_2$,  product of average single particle efficiencies, $\epsilon_1\epsilon_1=f_1f_1$, and the ratio $\epsilon_2/\epsilon_1\epsilon_1 =f_2/f_1f_1$ computed with the z vertex dependent efficiency model described in the text.}
\label{fig:EpsVsEtaVarZ}
\end{figure}
\end{widetext}
The ratio $f_2/f_1f_1$ may deviate significantly from unity and consequently biases the correlation function $R_2$.
 
\begin{figure}[h]
\includegraphics[width=0.45\textwidth]{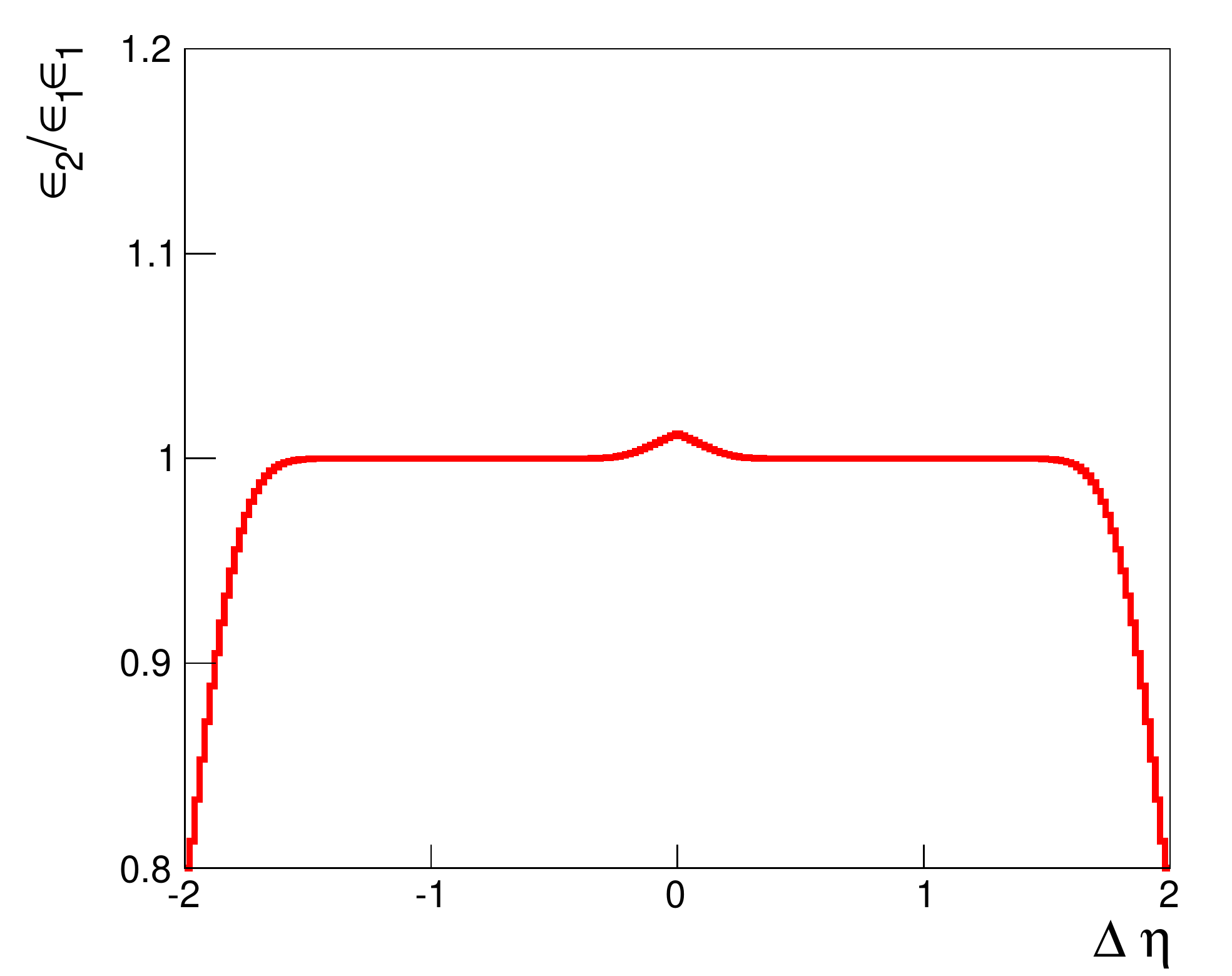}
\caption{(Color Online) Ratio $\epsilon_2/\epsilon_1\epsilon_1(\Delta\eta)$ computed with the efficiency model dependence on the z vertex described in the text.}
\label{fig:Eps2ToEps1Eps1VsDetaDirVarZ}
\end{figure}

We illustrate  this bias by considering experimental conditions such that the detection efficiency is function of the particle pseudorapidity and 
the vertex position. We modify our model of the $\eta$ dependence, Eq. \ref{eq:epsilonVsEta}, such that the "edges" $\eta_{<}$ and $\eta_{>}$
are  explicit functions of the primary vertex position.
\begin{eqnarray} 
\eta_{<} &= &-\eta_o - \alpha z \\ \nonumber
\eta_{<} &=&  \eta_o - \alpha z \\ \nonumber
\end{eqnarray} 
The coefficient $\alpha$ determines how fast the "edges" $\eta_{<} $ and $\eta_{>} $ shift(walk) with vertex position. 
We further assume, for illustrative purposes, that the vertex position is distributed according to a Gaussian distribution, $P_{c}(z)$, centered at $z=0$
and of width $\sigma_{col}$.
\begin{eqnarray} 
P_{c} = \frac{1}{\sqrt{2\pi}\sigma_{col}}e^{-z^2/2\sigma_{col}^2}
\end{eqnarray} 
The function $\epsilon(\eta|z)$ is shown in Fig. \ref{fig:Epsilon1VsEtaVsZ} for values $\alpha=0.02$,  $\sigma_{\epsilon}=0.1$ , and $\sigma_{col}=20$ cm for selected values of the vertex position, $z$. The pair efficiency average $\epsilon_2=f_2$, the product of average single particle efficiencies, $\epsilon_1\epsilon_1 = f_1f_1$, and the ratio $f_2/f_1f_1$ are plotted in Fig. \ref{fig:EpsVsEtaVarZ}. One notes that while the functions $f_2$ and $f_1f_1$ are flat over most of the detector acceptance, they exhibit rapid fall-off behavior near the edges 
of the acceptance. Given the finite width of the vertex distribution, the two functions   feature different rates and shapes of roll-off behavior. As illustrated in Fig. \ref{fig:EpsVsEtaVarZ} (c), one finds the ratio $f_2/f_1f_1$ therefore deviates significantly from unity towards the edges of the acceptance in $\eta_1, \eta_2$ space. This can lead to significant effects on $R_2(\Delta \eta)$. Assuming 
the correlation function is in fact equal to unity, i.e. $\la N_2(\eta_1,\eta_2) \ra/\la N_1(\eta_1) \ra \la N_1(\eta_2) \ra =1$, averaging the correlation function over $\overline{\eta}$ in the range $-1 < \eta <1$ leads to non trivial and significant deviations from unity towards the edges of the acceptance as well as for values near $\Delta \eta \sim 0$ as illustrated in Fig. \ref{fig:Eps2ToEps1Eps1VsDetaDirVarZ}. The magnitude and width of these deviations depend on the rolloff rate of the efficiency near the edge of the acceptance, the walk parameter $\alpha$, and the width of the vertex position distribution, $\sigma_{col}$. The magnitude of 
the walk parameter $\alpha$ determines the degree to which the factorization $\epsilon_1(\eta_1)\epsilon_1(\eta_2)$ is violated: the larger the spread, the less robust the observable $R_2$ becomes, irrespective of whether Method 1 or 2 are applied to determine it.

The robustness of the $R_2(\eta_1,\eta_2)$ observable can however be recovered if one changes the order in which the average over z and the ratio are taken. This can be accomplished by measuring the number of singles and pairs explicitly as a function of the vertex position $z$. The number of pairs and singles have to be binned in z. The appropriate number of bins must however be determined experimentally - as discussed below.

\begin{eqnarray} 
\label{Eq:etazR2}
R_2(\eta_1,\eta_2) &=& K \int_{z_{min}}^{z_{max}} P_c(z) \\ \nonumber
 & &\times \frac{\la n_2(\eta_1,\eta_2|z)\ra}{\la n_1(\eta_1|z) \ra\la n_1(\eta_2|z) \ra} dz
\end{eqnarray} 
The average number of singles and pairs being explicit functions of $z$, one can now write
\begin{eqnarray} 
R_2(\eta_1,\eta_2) &=&  K \int_{z_{min}}^{z_{max}} P_c(z) \\ \nonumber
 & &\times \frac{ \epsilon(\eta_1|z) \times \epsilon(\eta_2|z) \la N_2(\eta_1,\eta_2)\ra}{\epsilon(\eta_1|z) \times \epsilon(\eta_2|z)\la N_1(\eta_1) \ra\la N_1(\eta_2) \ra} dz
\end{eqnarray} 
provided the $z$-bins are sufficiently narrow to insure that the efficiencies do not appreciably fluctuate throughout a bin. The efficiencies thus once again cancel out, and one obtains the desired result.
\begin{eqnarray} 
R_2(\eta_1,\eta_2) &=&  K \frac{\la N_2(\eta_1,\eta_2)\ra}{\la N_1(\eta_1) \ra\la N_1(\eta_2) \ra}\int P_c(z)  dz \\ \nonumber&= &  \frac{\la N_2(\eta_1,\eta_2)\ra}{\la N_1(\eta_1) \ra\la N_1(\eta_2) \ra}
\end{eqnarray} 

\begin{figure}[h]
\includegraphics[width=0.48\textwidth]{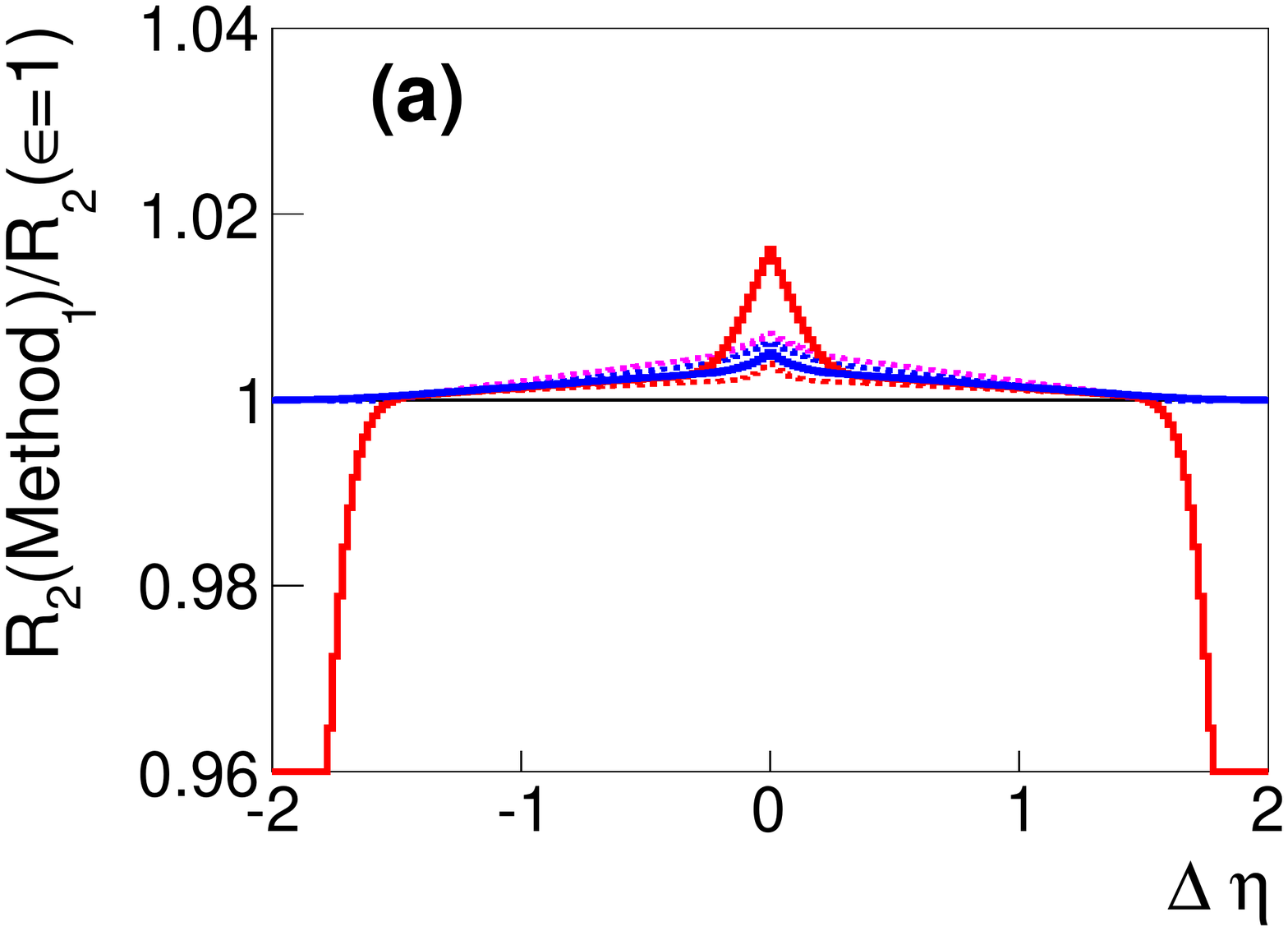}
\includegraphics[width=0.48\textwidth]{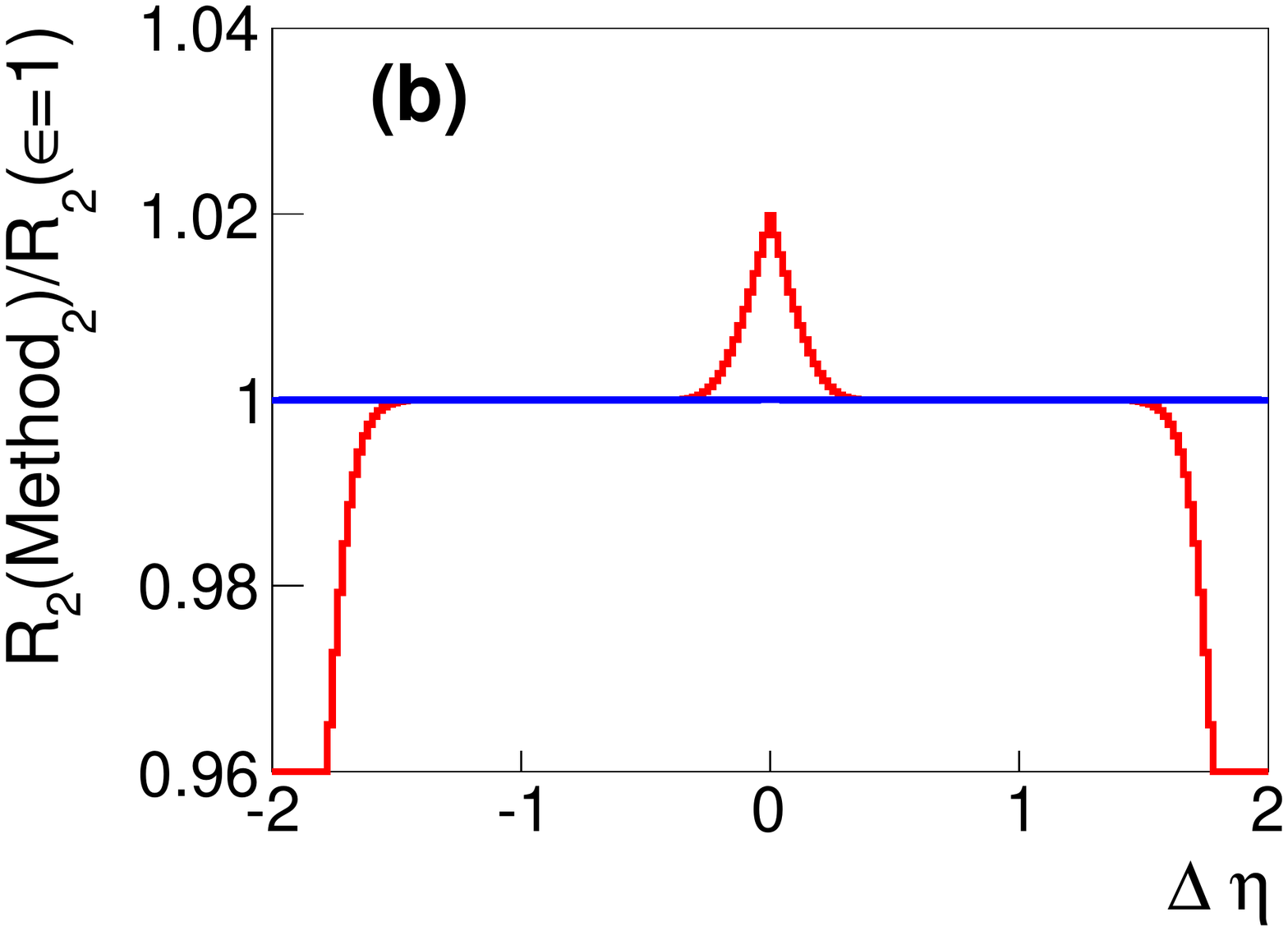}
\caption{(Color Online) Ratio of the $R_2$ correlation function reconstructed with (a) Method 1 and (b) Method 2 for several vertex position values and intervals as described in the text.  }
\label{fig:R2CompsVarZ}
\end{figure}

\begin{figure}[h]
\includegraphics[width=0.48\textwidth]{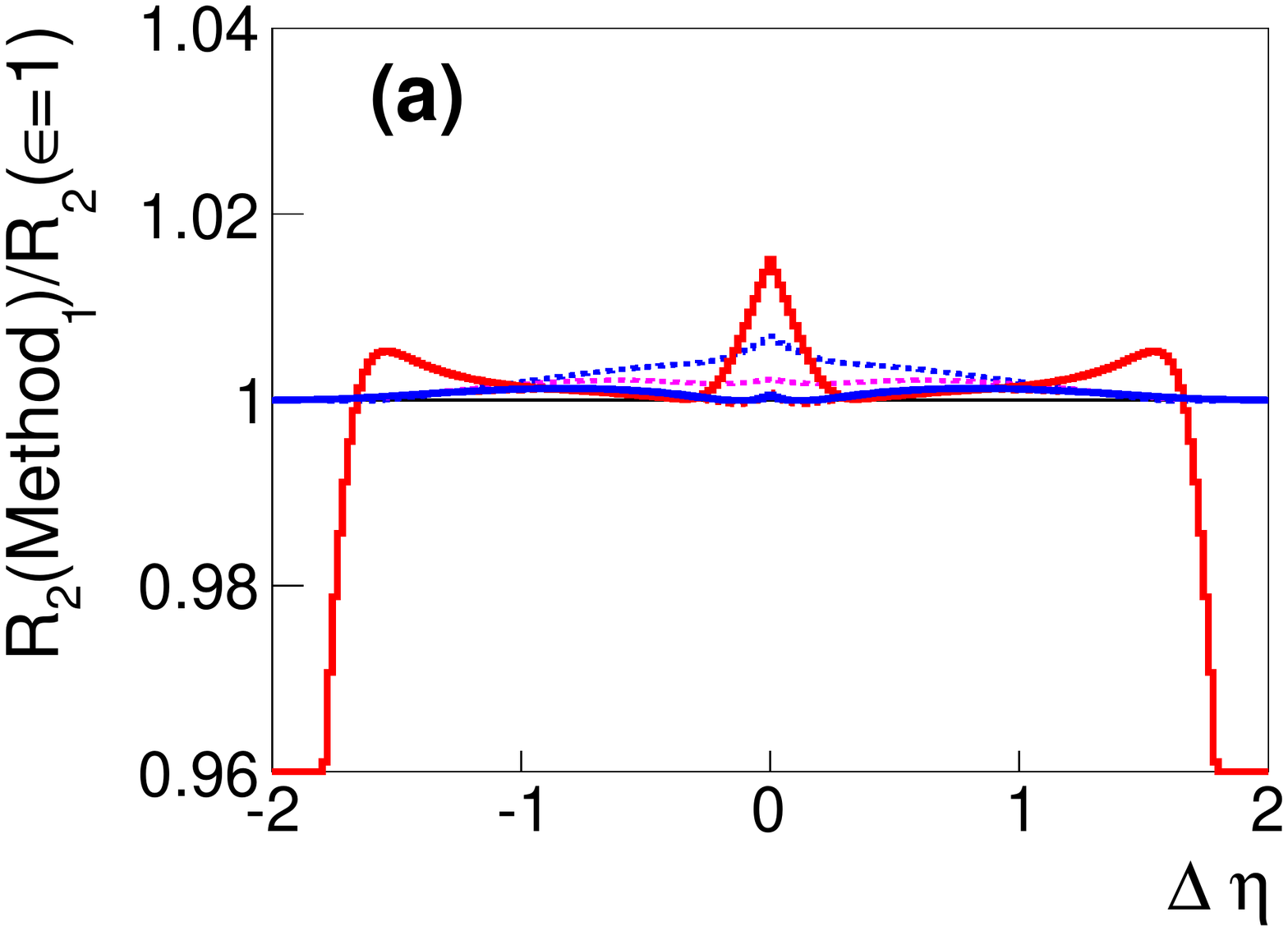}
\includegraphics[width=0.48\textwidth]{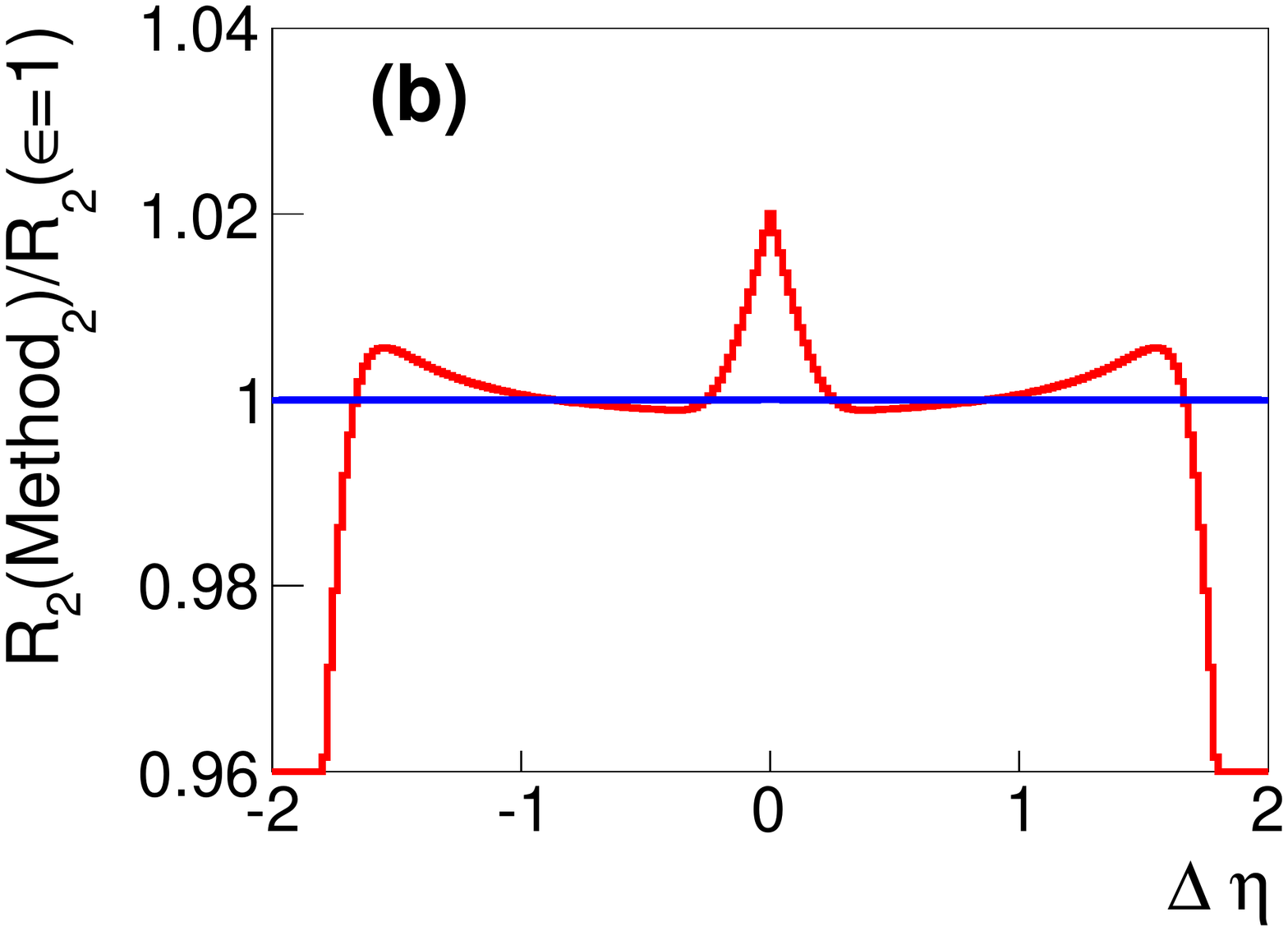}
\caption{(Color Online) Ratio of the $R_2$ correlation function reconstructed with (a) Method 1 and (b) Method 2 for a case where the efficiency exhibits a quadratic
dependence on $\eta$ as well as on the vertex position. The solid red curve corresponds to deviations obtained for $|z|<10 $ cm while the blue curve is obtained for $z$ bins of 0.25 cm width. Dash curves  correspond to various $z$ positions and intervals.  }
\label{fig:R2CompsVarZQuad}
\end{figure}

The cancellation of efficiencies achieved with Method 2 is exemplified in Fig. \ref{fig:R2CompsVarZ} (b) showing ratios 
of "measured" correlation functions $R_2$ by the correct correlation function (i.e. obtained with $\epsilon=1$). The red curve corresponds to 
an uncorrected  correlation function obtained within the range $|z|<10$ cm whereas the blue curve is obtained for $z$ bins of 0.25 cm. In the context of our model, such bins are sufficient to obtain virtually perfect (i.e. robust) correlation functions. By contrast, Fig.  \ref{fig:R2CompsVarZ} (a)  displays ratios for correlation function obtained with Method 1. Here again, the red curve exemplifies deviation 
obtained with no correction. The dash curves are obtained with various specific values of $z$ or $z$ intervals. All curves deviate from 
a perfect determination of the correlation function (solid black line).  Method 1 is simply {\it not} robust.
With Method 2, the size of the z-bin determines whether cancellation of the efficiencies properly take place. If the bins are too coarse, efficiency correlations persist and the observable is not robust. One finds that in the context of the model used above, a bin size of 0.5 cm leads to deviation of order 1 part per mil. Experimentally, if the the bins are too narrow, one may end up having too few events in a given bin thereby leading to numerical fluctuations or even infinities in the calculation of the ratio $\la n_2(\eta_1,\eta_2|z)\ra / \la n_1(\eta_1|z) \ra\la n_1(\eta_2|z) \ra$. The binning likely ends up being a compromise based on the size of the dataset and the level of precision sought after. 

Detector effects can further impact the amplitude and shape of correlation functions if the efficiency varies significantly through the acceptance and most particularly if such variations depend on the vertex position.  As an example, we here consider correlation functions obtained with the quadratic efficiency dependence  introduced in the sec. \ref{sec:efficiencyVsEta}.  As in the previous example, the response is  assumed to "walk" with the collision vertex position. Figure \ref{fig:R2CompsVarZQuad} presents ratios of the reconstructed correlation functions to the actual correlation function 
obtained with Methods 1 and 2. One finds once again that Method 2, as described in this section, enables to properly correct for both the efficiency and the dependence on the beam vertex position while correlation functions obtained with Method 1 belies the actual correlation function.

In closing, it is important to note that the technique described in this section enables measurements of correlation functions even when the efficiency is a strong function of the event vertex position as well as the particle rapidity. It thus makes it possible to stretch the fiducial acceptance of the measurement achievable with "traditional" and more conservative cut techniques. We emphasize once again that Method 1 is not robust in this context while Method 2 provides for a straightforward correction of instrumental effects.
 
\section{Collision centrality averaging}
\label{sec:collisionCentralityAveraging}

Correlation functions measured in heavy ion collisions are typically studied as a function of collision centrality determined with some global observables such as the total charged particle multiplicity and the transverse energy detected in some nominal acceptance providing an estimate
of the number of collision participants, or the energy
produced at zero degrees which is deemed proportional to the number of collision spectators.
We first discuss that a correction factor must be applied to account for the finite width of collision centrality bins
used in measurements of correlation functions, and subsequently show that the factorability of pair efficiencies can also 
break down in this context.

Correlation functions measured in heavy ion collisions are significantly different than those measured in  proton-proton interactions
and cannot be readily modeled with Eq. \ref{eq:decomposition}. Indeed, one finds that both the single and pair yields are functions
of collision centrality, $b$, which we denote as $n_1(\eta|b)$ and $n_2(\eta_1, \eta_2|b)$ respectively. 
Given sufficient statistics and computing resources, it should be possible to determine the correlation function $R_2$ as a function of $b$ with 
arbitrarily fine granularity (resolution) using method 2.
\begin{eqnarray} 
R_2(\eta_1, \eta_2|b) = \frac{n_2(\eta_1, \eta_2|b)}{n_1(\eta_1|b)n_1(\eta_2|b)}
\end{eqnarray}  
An average can then be taken over collision centralities in a specific range $[b_{min},b_{max}]$.
\begin{widetext}
\begin{eqnarray} 
\label{eq:r2Vsb}
R_2(\eta_1, \eta_2;b_{min}\le b <b_{max}) = \frac{\int_{b_{min}}^{b_{max}}   R_2(\eta_1, \eta_2|b) \sigma(b) db}{\int_{b_{min}}^{b_{max}}  \sigma(b) db}
\end{eqnarray} 
\end{widetext}
where $\sigma(b)$ is the collision cross section at centrality $b$.
In practice, it may not be possible to measure the single and pair densities with fine granularity. One
then gets densities as averages over collision centrality between the limits, $b_{min}$, and $b_{max}$.
\begin{widetext}
\begin{eqnarray} 
n_1(\eta;b_{min}\le b <b_{max}) &=& \frac{\int_{b_{min}}^{b_{max}} \rho_1(\eta|b) \sigma(b)db}{\int_{b_{min}}^{b_{max}}  \sigma(b)db} \\
n_2(\eta_1,\eta_2;b_{min}\le b <b_{max}) &=& \frac{\int_{b_{min}}^{b_{max}} \rho_2(\eta_1,\eta_2|b) \sigma(b) db}{\int_{b_{min}}^{b_{max}}  \sigma(b) db} \\ \nonumber
\end{eqnarray} 

An estimate of the correlation function (Eq.~\ref{eq:r2Vsb}) can  then be obtained from the ratio of the averages.
\begin{eqnarray} 
R_{2,{\rm est}}(\eta_1, \eta_2;b_{min}\le b <b_{max}) = \frac{n_2(\eta_1,\eta_2;b_{min}\le b <b_{max})}{n_1(\eta_1;b_{min}\le b <b_{max})n_1(\eta_2;b_{min}\le b <b_{max})}
\end{eqnarray} 
\end{widetext}
However note that this estimate exhibits a quadratic dependence on the bin width which must be corrected for. To demonstrate this dependence,
let us approximate the densities as follows:
\begin{eqnarray} 
n_1(\eta_1|b) =  n(b)  h_1(\eta_1) \\
n_2(\eta_1,\eta_2|b) =  n(b)(n(b)-1)  h_2(\eta_1,\eta_2) \\ \nonumber
\end{eqnarray}  
where the average integrated single yield, $ n(b)$, and the average pair yield, $n(b)(n(b)-1)$, both evaluated at a fixed value of $b$, are assumed to
embody the centrality dependence while the functions $h_1(\eta_1)$ and $h_2(\eta_1,\eta_2)$ are assumed
independent of collision centrality. The above estimate of the correlation function then becomes
\begin{widetext}
\begin{eqnarray} 
\label{eq:r2VsbEst}
R_{2,{\rm est}}(\eta_1, \eta_2;b_{min}\le b <b_{max}) =  Q(b_{min},b_{max})\frac{h_2(\eta_1,\eta_2)}{h_1(\eta_1)h_1(\eta_2)}
\end{eqnarray} 
with
\begin{eqnarray} 
Q(b_{min},b_{max}) = \frac{\int_{b_{min}}^{b_{max}} n(b)\left(n(b)-1\right) \sigma(b)db}{ \left(\int_{b_{min}}^{b_{max}} n(b) \sigma(b)db\right)^2}> 1
\end{eqnarray} 
which is manifestly different than the result obtained with Eq.~\ref{eq:r2Vsb} for which one has:
\end{widetext}
\begin{eqnarray} 
\frac{\int_{b_{min}}^{b_{max}} \frac{n(b)\left(n(b)-1\right)}{n(b)n(b)} \sigma(b)db}{\int_{b_{min}}^{b_{max}}\sigma(b)db } \approx 1
\end{eqnarray} 
The estimate Eq.~\ref{eq:r2VsbEst} is consequently increasingly biased with increasing bin width.  This
effect may however be suppressed  by dividing $R_{2,{\rm est}}$ explicitly by  $Q(b_{min},b_{max})$.

This correction may unfortunately become insufficient if the detection efficiency exhibits a dependency on the detector occupancy and
hence on collision centrality, $b$. Writing the detection efficiency as an explicit function of $b$, one finds 
\begin{widetext}
\begin{eqnarray} 
n_1(\eta;b_{min}\le b <b_{max}) &=& \frac{\int_{b_{min}}^{b_{max}}\epsilon_1(\eta|b) \rho_1(\eta|b) \sigma(b)db}{\int_{b_{min}}^{b_{max}}  \sigma(b)db} \\
n_2(\eta_1,\eta_2;b_{min}\le b <b_{max}) &=& \frac{\int_{b_{min}}^{b_{max}}\epsilon_1(\eta_1|b)\epsilon_1(\eta_2|b) \rho_2(\eta_1,\eta_2|b) \sigma(b) db}{\int_{b_{min}}^{b_{max}}  \sigma(b) db} \\ \nonumber
\end{eqnarray} 
\end{widetext}
The estimate Eq.~\ref{eq:r2VsbEst} is thus not only biased due the width of the centrality bin but also intrinsically non robust against detection efficiencies. To obtain a robust quantity, one must reverse the order in which the ratio of pair yields to product of single particle yields and the averaging over impact parameter $b$ are taken. This is accomplished by first calculating
the estimate Eq.~\ref{eq:r2VsbEst} using narrow bins in centrality. The ratios $R_2(\eta_1, \eta_2|b)$ obtained in narrow impact parameter bins 
must next be averaged across the centrality bin with Eq.~\ref{eq:r2Vsb}. In principle, this calculation technique
yields properly corrected, and robust, correlation functions in the limit of very narrow centrality bins. Its feasibility
may however be limited by the size of the data sample. If too fine a binning in $b$ is attempted, the sampled single particle yield 
$\rho_1(\eta|b)$ may be null in one or several $\eta$ bins. The ratio $R_2(\eta_1, \eta_2|b)$ would then 
diverge in those $\eta$ bins, and the method would consequently fail. This implies that, in practice, it is necessary
to systematically test the correction method and  verify its convergence while changing the impact parameter bin size
used to carry out the corrections.

\section{Efficiency dependence on the instantaneous collider luminosity and detector occupancy }
\label{sec:efficiencyVsL}

At colliders, the instantaneous luminosity and collision rates  varies considerably between the beginning and end of a fill. The detector hit occupancy is thus subject to change, sometimes dramatically, through a "run". The charged particle track detection and reconstruction efficiency unfortunately varies  inversely  to the occupancy. The single and pair yields may thus be a function of time and/or instantaneous luminosity. Adding to this effect, note that in a drift detector such as a TPC, the space charge load may affect differently the tracks detected at short and long drift times thereby leading to complicated efficiency dependence on pseudo rapidity, vertex position, and instantaneous luminosity. This may be particularly important for analyses of correlations between identified particles e.g. kaons, pions, etc. Particle identification efficiency requires one puts cuts on specific energy loss ($dE/dx$) or time of flight observables which may be affected by varying levels of detector occupancy. Kaon and pion yields may then be changing with the luminosity \cite{westfall}. Detection efficiencies thus become explicit functions of the instantaneous luminosity, $L$, delivered to the detector, $\epsilon(\eta_i| L)$. As for dependencies on the vertex position, it is straightforward to verify that efficiencies do not cancel out in the measurement of $R_2(\eta_1,\eta_2)$. Robustness can however be recovered, as in the case of vertex position dependent efficiencies, if the measured number of single and pairs can also been determined according to the estimated instantaneous luminosity (or detector occupancy). Expressing the fraction of events measured with a given luminosity $L$ as $P_{\rm Lum}(L)$, and binning the measurements of singles and pairs with sufficient granularity in $L$, the correlation function $R_2$ can thus formally be written:
\begin{widetext}
\begin{eqnarray} 
\label{Eq:LumoR2}
R_2(\eta_1,\eta_2) = K' \int_{L_{min}}^{L_{max}} P_{\rm Lum}(L) \frac{\la n_2(\eta_1,\eta_2|L)\ra}{\la n_1(\eta_1|L) \ra\la n_1(\eta_2|L) \ra} dL
\end{eqnarray} 
\end{widetext}
where $K'$ is a normalization constant.
\begin{eqnarray} 
K'^{-1} = \int_{L_{min}}^{L_{max}} P_{\rm Lum}(L) dL
\end{eqnarray} 
Eq. \ref{Eq:LumoR2} is structurally similar to Eq. \ref{Eq:etazR2}. The same analysis techniques can thus be used in cases where the efficiency is a complex function of some external but controllable parameter or agent. One can obviously extend either analyses by considering joint dependencies on $z$ and $L$ explicitly, or any other external variables affecting the detection efficiency. Once again, the use of Method 2 enables proper correction of the $R_2$ observable while Method 1 may introduce various biases because $\overline{\eta}$ averaging eliminates the factorability of the pair efficiency.

\section{Weight technique}
\label{sec:efficiencyVsPhi}

No detector provides perfect coverage. Though existing Time Projection Chambers (TPC) such as the STAR and ALICE TPCs come close to a perfect azimuthal coverage, in collider geometry, there effectively remains some small gaps between their readout sectors. Sensors have varied performances. One ends up having detection efficiencies that  depend on the particle azimuth of emission, $\phi$. Techniques based on mixed events, or measurements of singles, are routinely used to account for these non uniformities and are for the most part very successful in eliminating glitches at the boundaries between sectors in two-particle correlation studies carried out as a function of the difference of azimuth, $\Delta \phi = \phi_1 - \phi_2$. There is however a class of observables where additional steps have to be taken to suppress sector boundary effects.  Two-particle correlation functions can be  generalized to study particle production dynamics dependencies on transverse momentum or other variables of interest. Integral transverse momentum correlation functions, $\la \Delta p_{\rm T} \Delta p_{\rm T}\ra$, are of interest, in particular, as they enable integral measurements of momentum correlation that are sensitive to temperature fluctuations and might be useful, at least in principle, in identifying the position of the tri-critical point of nuclear matter \cite{stephanov1999}. Integral transverse momentum correlation functions can also 
be generalized into differential $\Delta p_{\rm T} \Delta p_{\rm T}$ weighted differential correlation function as follows \cite{pruneauDptDpt}:
\begin{widetext}
\begin{eqnarray} 
\la \Delta p_{\rm T} \Delta p_{\rm T} (\Delta \eta, \Delta \phi)\ra = \frac{\int \rho_2(\vec{p}_1,\vec{p}_2) \Delta p_{{\rm T},1}\Delta p_{{\rm T},2} dp_{{\rm T},1}dp_{{\rm T},2}}
{\int \rho_2(\vec{p}_1,\vec{p}_2) dp_{{\rm T},1} dp_{{\rm T},2}}
\end{eqnarray} 
\end{widetext}
where $\Delta p_{T,i}=p_{T,i} - \la p_{\rm T}\ra$, with $i=1,2$ measures the $p_{\rm T}$  deviation to the mean transverse momentum, $\la p_{\rm T}\ra$, for each produced particle. This type of differential transverse momentum correlation adds explicit sensitivity to momentum correlations. Note in particular that the quantity $\Delta p_{{\rm T},1}\Delta p_{{\rm T},2}$ is not 
positive definite. A pair consisting of two particles above or below the mean momentum will yield a positive value but one below and one above the mean will yield a negative value for this coefficient. The introduction of this coefficient consequently enables further sensitivity to the correlation dynamics, i.e. whether particle correlations are dominated by particle pairs yielding a positive or negative $\Delta p_{{\rm T},1}\Delta p_{{\rm T},2}$ coefficient. The correlation function is also sensitive to flow effects, and provides a test of the factorability of flow coefficients\cite{pruneauDptDpt}. This type of correlation function however introduces new challenges in data analyses because of the explicit and manifest dependence of the correlation function on $p_{\rm T}$  and its dependence on the azimuthal angles of particle emission, $\phi_i$.  TPC sector boundaries, in particular, introduce the need for additional corrections because not only is the efficiency a function of the azimuth, it is also a function of the transverse momentum of the particles. In general, the two dependencies do not factorize. Even worst, the dependence on $p_{\rm T}$  may be found to be a rather intricate function of $\phi$. This poses an important problem because the correlation function includes an explicit dependence on $p_{\rm T}$ . Effectively, the uncorrected average transverse momentum becomes a complicated function of $\phi$ because of the intricate dependence of the efficiency on both $\phi$ and $p_{\rm T}$ . In fact, one finds that the $p_{\rm T}$  spectrum of the produced particle also acquires intricate dependencies on $\phi$, all because of instrumental effects. Given the effects are associated with sector boundaries, one ends up getting artificial modulations  in $\Delta \phi$ in measurements of $\la \Delta p_{\rm T} \Delta p_{\rm T} (\Delta \eta, \Delta \phi)\ra$. To remedy these instrumental effects, one must equalize the response simultaneously in $\phi$ and $p_{\rm T}$. The correction technique in terms of mixed events or product of singles described in previous sections can in principle be
utilized to achieve this goal but it becomes prohibitively expensive in terms of memory and required statistics. It therefore ends up being somewhat impractical. An alternative technique, which we recommend, instead, is to introduce instrumental weights, $\omega(\eta,\phi,p_{\rm T})$ in the determination of  correlation  functions. 

The weight technique is rather general and may be used towards the study of $R_2$ as well as $\la \Delta p_{\rm T} \Delta p_{\rm T} (\Delta \eta, \Delta \phi)\ra$. The weights are designed to equalize the detector response. A technique to obtain them is discussed below. Once the weights are available, one can proceed to carry out the correlation analysis using the technique presented in the prior sections and by incrementing histograms with weights 
$\omega(\eta_1,\phi_1,p_{{\rm T},1}) \omega(\eta_2,\phi_2,p_{{\rm T},2})$ rather than unity, for $R_2$ analyses, and $\omega(\eta_1,\phi_1,p_{{\rm T},1}) \omega(\eta_2,\phi_2,p_{{\rm T},2})\Delta p_{{\rm T},1}\Delta p_{{\rm T},2}$ for $\la \Delta p_{\rm T} \Delta p_{\rm T}\ra$ analyses. 

In the context of $\Delta\eta$ vs. $\Delta\phi$ analyses, one must account for the fact that detection efficiencies are complicated functions of $\phi$, $\eta$, and $p_{\rm T}$  that may evolve with the position, $z$, of the collision vertex. One must consequently obtain weights, $\omega(\eta,\phi,p_{\rm T},z)$, that depend simultaneously on all four of these variables.  The weights thus acquire a dual function: they account for the $z$ dependence  as well as  the $p_{\rm T}$  vs. $\phi$ dependencies simultaneously. The purpose  of the weights is to equalize the response in $p_{\rm T}$  across all values of $\phi$, $\eta$, and for all $z$. They can thus be calculated as
\begin{eqnarray} 
\omega(\eta,\phi,p_{\rm T},z) = \frac{\int d\phi \int d\eta \int_{z_{min}}^{z_{max}}dz   \la n(\eta,\phi,p_{\rm T},z) \ra  }{\la n(\eta,\phi,p_{\rm T},z) \ra}
\end{eqnarray} 
where the integration on $\phi$ and $\eta$ covers the fiducial acceptance of interest. By construction, the $p_{\rm T}$  spectra and $\la p_{\rm T}\ra$ become independent of $\phi$ as well as $\eta$.  Independence relative to $\eta$ is likely acceptable  at LHC and RHIC in the context of narrow $\eta$ acceptance detectors such as STAR and ALICE but better $\eta$ dependent treatment may be required for wider acceptances. 
\bigskip
\section{Summary}
\label{sec:summary}

We have compared the merits of two methods commonly used to correct for effects associated with limited and non-uniform detection efficiencies.  
Method 1 involves a ratio of 1D-histograms incremented as a function of $\Delta\eta$ for same and mixed events pairs of particles while Method 2 proceeds on the basis of a ratio of 2D-histograms, measured  as a function of $\eta_1$ and $\eta_2$, similarly obtained from same and mixed events pairs of particles. One can also use a product of single particle yields, $\rho_1(\eta_1)\rho_1(\eta_2)$, in lieu of mixed events. Method 2 calculates 
the correlation function $R_2(\Delta\eta)$ by explicitly averaging the ratio of 2D-histograms over $\overline{\eta}$. 
We showed that Method 2 is intrinsically robust against variations of the efficiency across the detector fiducial acceptance provided 
the pair efficiency factorizes, $\epsilon_2=\epsilon_1\epsilon_1$, while Method 1 is not robust and exhibits finite dependencies on the efficiencies. We also showed that efficiency dependencies on "external" parameters such as the vertex position or the instantaneous luminosity break down the assumption of the factorability of the pair efficiency and render correlation functions obtained with both Methods non-robust. Robustness can however be recovered in the case of Method 2 by studying the correlation function in narrow bins of $z$-vertex, beam luminosity, or detector occupancy.
Given the determination of $R_2$ with fine binning in z-vertex position (or other global event variables that break the factorability of the pair efficiency) 
may be computationally tedious, we introduced a weight method that account for the varied detector responses with kinematic and global event variables based on a single set of histograms.

\section*{Acknowledgement}

The authors wish to thank Michael Weber, Constantin Loizides, and Sergei Voloshin for their critical reading of the manuscript and the constructive comments they provided. This work was supported in part by DoE grant DE-FG02-92ER-40713.



\end{document}